\pdfoutput=1
\documentclass[12pt]{article} 
\usepackage{geometry}
\usepackage{amsmath} % already in sashorthand but should be defined earlier here to use \numberwithin
\numberwithin{equation}{section}
\RequirePackage[pdftex,dvipsnames]{xcolor}
\definecolor{darkblue}{rgb}{0.1,0.1,.7}
\usepackage[colorlinks, linkcolor=darkblue, citecolor=darkblue, urlcolor=darkblue, linktocpage]{hyperref} 

\usepackage[square, comma, sort&compress,numbers]{natbib} % reference management

\usepackage[margin=10pt,font=small,labelfont=bf]{caption}

\usepackage{tikz}
\usetikzlibrary{external}

\usepackage[]{sashorthand} % Provides useful commands and packages. It has as three potions:
\usepackage[final]{pdfpages}

\DeclareMathAlphabet{\mathpzc}{OT1}{pzc}{m}{it} % New font \mathpzc
\begin{document}
	\vspace*{-.6in} \thispagestyle{empty}
	\begin{flushright}
	\end{flushright}
	\vspace{.2in} {\Large
		\begin{center}
			{\bf An \'etude of momentum space scalar amplitudes in AdS\vspace{.1in}}
		\end{center}
	}
	\vspace{.2in}
	\begin{center}
		{\bf 
			Soner Albayrak$^{a,b}$, Chandramouli Chowdhury$^{c}$, and Savan Kharel$^{d}$}
		\\
		\vspace{.2in} 
		${a}$ {\it  Department of Physics, Yale University, New Haven, CT 06511}\\
		${b}$ {\it  Walter Burke Institute for Theoretical Physics, Caltech, Pasadena, CA 91125}\\
		${c}$ {\it International Centre for Theoretical Sciences, Tata Institute of Fundamental Research, Sivakote, Bangalore 560089, INDIA}\\
		%$^c$ {\it ICTS, Tata Institute of Fundamental Research, Sivakote, Bangalore 560089, INDIA}\\
		$
		{d}$ {\it  Department of Physics, Williams College, Williamstown, MA 01267}
	\end{center}
	
	\vspace{.2in}
	
\begin{abstract}
In this paper, we explore momentum space approach to computing scalar amplitudes in Anti-de Sitter space. We show that the algorithm derived by
Arkani-Hamed, Benincasa, and Postnikov for cosmological wavefunctions can be straightforwardly adopted for AdS transition amplitudes in momentum space, allowing one to bypass bulk point integrations. We demonstrate the utility of this approach in AdS by presenting several explicit results both at tree and loop level.
\end{abstract}
	
	\newpage
	
	\tableofcontents
	
%	\newpage
	
% BODY
\section{Introduction}
In the last decade, there has been a resurrection in the study of scattering amplitudes and conformal correlation functions. Such undertakings have extricated rich structures of quantum field theory and quantum gravity. In particular, we now have considerable evidence that scattering amplitudes in quantum gravity can be computed from the correlation functions in one lower dimensions. Such a correspondence is known as the holographic duality  \cite{Maldacena:1997re, Witten:1998qj, Gubser:1998bc} and its most concrete formalism in given by AdS/CFT  where the bulk geometry is Anti de Sitter space and the conformal correlation functions live at the boundary. This correspondence has led to major insights into the nature of quantum gravity as well as gauge theory.

Besides yielding useful insights, holographic correlators have been discovered to have rich mathematical structures \cite{Freedman:1998bj,Liu:1998ty,Freedman:1998tz,DHoker:1999mqo,DHoker:1999kzh,Penedones:2010ue,Paulos:2011ie,Mack:2009gy, Fitzpatrick:2011ia, Kharel:2013mka, Fitzpatrick:2011hu, Fitzpatrick:2011dm, Costa:2014kfa, Jepsen:2018ajn, Jepsen:2018dqp, Gubser:2018cha, Hijano:2015zsa,Yuan:2018qva, Ghosh:2018bgd, Liu:2018jhs, Parikh:2019ygo, Jepsen:2019svc, Penedones:2019tng, Aprile:2019rep, Ponomarev:2019ofr, Meltzer:2019pyl, Meltzer:2019nbs,Caron-Huot:2018kta,Fichet:2019hkg}. However, momentum-space methods,  our usual modus operandi of doing computation in quantum field theory, is still not yet fully studied for CFTs (see for a partial list of progress \cite{Raju:2010by, Mata:2012bx, Raju:2012zs, Raju:2012zr, Raju:2011mp, Arkani-Hamed:2015bza, Bzowski:2013sza, Bzowski:2015pba,Bzowski:2018fql, Bzowski:2015yxv, Isono:2019ihz, Isono:2018rrb,Isono:2019wex, Coriano:2018bbe, Coriano:2019dyc, Maglio:2019grh, Gillioz:2018mto, Arkani-Hamed:2018kmz, Coriano:2018bsy, Coriano:2013jba, Bzowski:2019kwd, Anand:2019lkt, Gillioz:2019lgs, Farrow:2018yni, Lipstein:2019mpu, Nagaraj:2019zmk}). We also now know that the usual scattering amplitudes arise from the flat space limit of the holographic Conformal Field Theory (CFT) correlators. Hence it is useful to generalize the tools that have been developed for the usual scattering amplitudes in flat space to holographic correlators. A modest step in this direction is taken in \cite{Albayrak:2018tam} where it is shown that tree level gauge theory Witten diagrams for transition amplitudes, reduce to surprisingly simple expressions when expressed  in momentum space.\footnote{Transition amplitudes are generalizations of vacuum-correlators such that one replaces some of the bulk to boundary propagators of the relevant Witten diagram with normalizable modes \cite{Raju:2011mp}. Such a replacement roughly creates the effect of changing the boundary conditions at past and future horizons of the relevant Poincar\'e patch, creating past and future states for the correlator (hence the name transition amplitude).} In a similar vein, momentum space approach to transition amplitudes also simplifies the computation of  graviton exchange diagrams, which was demonstrated in \cite{Albayrak:2019yve} with explicit higher point tree level results.

The momentum space formalism for AdS calculations can be upgraded with a new algorithm developed by Arkani-Hamed, Benincasa, and Postnikov in \cite{Arkani-Hamed:2017fdk} where they investigate the wavefunction of the Universe in de Sitter background. Indeed, it was realized in \cite{Albayrak:2019asr} that the computation of gluon Witten diagrams in AdS$_4$ can actually use the same combinatorial relations developed in \cite{Arkani-Hamed:2017fdk} if it is written in momentum space. This correspondence allowed the authors to compute any tree level gluon exchange diagram algebraically, without having to do any explicit bulk integrations.

In this paper, we would like to extend this marriage between Arkani-Hamed et al's algorithm and AdS momentum space beyond AdS$_4$ gluons. We will show that conformally coupled scalars in any AdS$_{d+1}$ can be computed algorithmically, both at tree and loop level, and we will demonstrate this with explicit results for various Witten diagrams. Besides its formal usage, the AdS transition amplitudes can be useful in the computation of  the wave function at late times from which one can compute de Sitter correlators \cite{Maldacena:2002vr,Ghosh:2014kba}. The growing interest in cosmology has generated a great deal of excitement in the study of late time de Sitter correlators   \cite{McFadden:2009fg, Chen:2009zp, Maldacena:2011nz, Baumann:2011nk, Assassi:2012zq,Chen:2012ge,Assassi:2013gxa,Arkani-Hamed:2015bza,Lee:2016vti,An:2017hlx,Kehagias:2017cym,Kumar:2017ecc,Baumann:2017jvh,Franciolini:2017ktv,Arkani-Hamed:2018kmz,Goon:2018fyu,Anninos:2019nib,Pi:2012gf,Gong:2013sma,Sleight:2019mgd,Sleight:2019hfp,Hillman:2019wgh} and we believe that the analogous calculations of momentum space AdS amplitudes can assist in the study of the shape of non-Gaussianities. 

Here is the organization of the paper. In \secref{\ref{sec: preliminaries}}, we discuss scalars in curved spacetime and present the review of momentum space toolkit in Anti-de Sitter space. We also demonstrate the standard non-algorithmic approach of momentum space formalism by computing Witten diagrams for minimally coupled scalars. In \secref{\ref{sec: conformally coupled scalars}}, we switch to conformally coupled scalars, discuss how the algorithmic approach works, and provide explicit results both at tree and loop levels. Finally, we conclude with a brief discussion and future directions.

\section{Preliminaries}
\label{sec: preliminaries}
\subsection{Scalars in curved spacetime}
\label{sec: scalars in curved spacetime}
A free scalar $\f$ in flat space satisfies the Klein-Gordon equation which reads as
\be 
\left(\square-m^2\right)\Phi=0
\ee 
where $m$ is the mass parameter of the field. In curved spacetime, this equation becomes
\be 
\label{eq: EOM}
\left(\square-(m^2+\xi R)\right)\Phi=0
\ee 
where $R$ is the Ricci scalar and $\xi$ is a coefficient determining the interaction between scalar and the background. In the case of AdS$_{d+1}$, this equation follows from the action
\be 
S_\text{quadratic}=-\half\int d^{d+1}x\sqrt{g}\left(g^{\mu\nu}(\partial_\mu\Phi)(\partial_\nu\Phi)+(m^2+\xi R)\Phi^2\right)
\ee 
where $g=\abs{\det g_{\mu\nu}}$ and we stick to mostly positive metric convention throughout the paper.

The scalar in curved spacetime has been extensively analyzed in the literature; however, the analysis usually focus on two specific values of $\xi$: \emph{minimally coupled scalar} with $\xi\rightarrow0$, and \emph{conformally coupled scalar} with $\xi\rightarrow \xi_c$ for 
\be 
\xi_c\equiv\frac{d-1}{4d}\;.
\ee
The popularity of minimally coupled scalar follows from the fact that $\xi=0$ simplifies the Lagrangian. The appeal of conformally-coupled scalar, however, cannot be immediately seen unless one goes to local Minkowski frame where the potential term takes the form \cite{Sonego:1993fw}
\be 
V(x)\sim\left[m^2+(\xi-\xi_c)R\right]
\ee 
One sees that if (in addition to $\xi=\xi_c$) one imposes $m=0$, the potential vanishes, leading the theory to enjoy conformal symmetry. Indeed, even though all pairs $(m,\xi)=(m,\xi_c)$ fall into the class of \emph{conformally coupled scalars}, only $(m,\xi)=(0,\xi_c)$ case is invariant under conformal transformations \cite{Birrell:1982ix}.

Tuning the parameters $(m,\xi)=(0,\xi_c)$ is necessary for the theory to enjoy conformal symmetry but it is not sufficient: we also need to check if the interaction Lagrangian spoils this symmetry. As a prerequisite condition of scale invariance we restrict to interactions of the form $\cO\equiv\cO(g^{\mu\nu},\partial_\mu,\Phi)$ which transforms as $\cO\rightarrow \lambda^\kappa\cO$ as $g_{\mu\nu}\rightarrow\lambda^2g_{\mu\nu}$ for constant $\lambda$.

In this paper, we focus on non-derivative interactions for which the action takes the form
\be 
\label{eq: polynomial lagrangian}
S=-\int d^{d+1}x\sqrt{g}\left[\half\left(g^{\mu\nu}(\partial_\mu\Phi)(\partial_\nu\Phi)+(m^2+\xi R)\Phi^2\right)+\frac{\lambda_n}{n!} \Phi^n\right]
\ee 
We can check the trace of stress tensor $T^{\mu\nu}$ to see when it is zero. Indeed, via
\be 
T_{\mu\nu}=\frac{-2}{\sqrt{g}}\frac{\delta S}{\delta g^{\mu\nu}}
\ee
we obtain
\begin{multline}
 T_{\mu\nu}=(\partial_\a\Phi)(\partial_\b\Phi)\left[\delta^\a_\mu\delta^\b_\nu-\half g_{\mu\nu}g^{\a\b}\right]-\xi\left[\partial_\mu\partial_\nu-g_{\mu\nu}\square\right]\Phi^2\\-\half\Phi^2\left[\left(m^2+\xi R\right)g_{\mu\nu}-2\xi R_{\mu\nu}\right]-\frac{\lambda_n}{n!} \Phi^n g_{\mu\nu}
\end{multline}
where we used
\be 
\delta F(R)=\left(F'(R)R_{\mu\nu}-\left[\nabla_\mu\nabla_\nu-g_{\mu\nu}\square\right]F'(R)\right)\delta g^{\mu\nu}\;.
\ee 
for whose derivation with a  nice explanation we suggest the lecture notes of Matthias Blau, available at \hyperref{http://www.blau.itp.unibe.ch/newlecturesGR.pdf}{}{}{http://www.blau.itp.unibe.ch/newlecturesGR.pdf}.\footnote{
This equation is actually true only if there are  no contributions at the boundary as its derivation uses integration by parts and assumes that total derivative terms do not contribute. Strictly speaking, this is not correct for AdS. However, boundary conditions actually kill the additional piece unless the variation of the action on the boundary contains the derivative of the variation of the boundary metric for which one then needs to add an appropriate boundary term to cancel the additional variation, see \cite{Gibbons:1976ue} in the case of Einstein gravity. We will avoid such subtleties (and refer reader to \cite{Guarnizo:2010xr,Nojiri:2000gv} and references therein) so a rigorous minded reader should see our calculations not as a derivation but as a motivation for why only certain interactions can enjoy full conformal symmetry.
}

The trace of stress tensor then reads as
\be 
g^{\mu\nu}T_{\mu\nu}=2 d (\xi-\xi_c)g^{\mu\nu}(\partial_\mu\Phi)(\partial_\nu\Phi)+2d\xi\Phi\left[\square\Phi-\left(\frac{d+1}{d-1}m^2+\xi_c R\right)\Phi-\frac{\lambda_n}{n!}\frac{\xi_c}{\xi}\frac{2(d+1)}{d-1}\Phi^{n-1}\right]
\ee 
We see the first term dies only if $\xi=\xi_c$. We can then kill the second term with equation of motion if $m=0$ and $n=n_c$ for
\be 
\label{eq: n_c}
n_c\equiv \frac{2(d+1)}{d-1},
\ee 
e.g. $n_c=4$ for AdS$_4$.
We thus arrived at the well-known conclusion: the action in \equref{eq: polynomial lagrangian} enjoys conformal symmetry only if $\{m,\xi,n\}=\{0,\xi_c,n_c\}$.

We can derive this result from another, and slightly simpler, approach. We first specialize to the AdS with the Poincar\'e metric and the Ricci scalar
\be 
\label{eq: poincare patch}
ds^2 =\frac{dz^2 +  \eta_{ij} dx^i dx^j}{(z/ \rho)^2}\;,\quad R=-\frac{d(d+1)}{\rho^2}
\ee 
where we take AdS radius $\rho=1$ in the rest of the paper.\footnote{Our notation is such that $z$ is the radial coordinate and the transverse coordinates $x_i$ approach to the boundary coordinate as $z\rightarrow 0$.} We then consider a Weyl transformation which maps AdS to flat space:
\bea[eq:Weyl transformation]
g_{\mu\nu}\quad\rightarrow&\quad g'_{\mu\nu}\equiv z^2g_{\mu\nu}\\
\Phi\quad\rightarrow&\quad \phi\equiv z^{-\frac{d-1}{2}}\Phi\label{eq: field weyl transformation}
\eea 
where we used  the engineering scaling dimension for the scalar field. Under this transformation, quadratic part which is invariant under conformal transformations map to the free scalar in flat space
\begin{subequations}
	\begin{equation}
	\half \sqrt{g}g^{\mu\nu}(\partial_\mu\Phi)(\partial_\nu\Phi)+\half \xi \sqrt{g}R\Phi^2 \quad\rightarrow\quad \half(\partial_\mu\phi)(\partial^\mu\phi)
	\end{equation}
	whereas the interaction part maps as 
	\begin{equation}
	\label{eq: flat space interaction}
	\sqrt{g} \frac{\lambda_n}{n!} \Phi^n \quad\rightarrow\quad z^{\frac{d-1}{2}\left(n-n_c\right)}\frac{\lambda_n}{n!}\phi^n
	\end{equation}
\end{subequations}
We immediately see that we need $n=n_c$ if we require the flat space interaction to be conformally invariant as well.

\subsection{Review of momentum space toolkit in AdS}
\label{sec: review of momentum space}
In this section, we will review the basics of our framework and specifics regarding the scalars. For similar reviews in the context of gauge fields and gravitons, see \cite{Albayrak:2018tam,Albayrak:2019asr,Albayrak:2019yve}.

We will be working with the Poincar\'e patch of \equref{eq: poincare patch} and take the Fourier transform of $x_i$. We will leave $z$ as it is though: the coordinates $\{z,k_i\}$ are what we call the \emph{momentum space} in this paper. This is in the same spirit of the treatment in \cite{Raju:2012zs,Raju:2011mp,Raju:2010by}.

In momentum space, equation of motion in \equref{eq: EOM} becomes
\be 
\left(z^{1+d}\partial_z z^{1-d}\partial_z-z^{2}k_ik^i-\mu^2\right)\Phi(z,k_i)=0
\ee 
where we define the \emph{effective mass square}
\be 
\mu^2\equiv m^2+\xi R
\ee 

The general solution to this differential equation for timelike momenta $k_ik^i<0$ reads as:\footnote{Here, $J_n(x)$ ($Y_n(x)$) is the Bessel function of the first (second) kind.}
\be 
\Phi(z)\sim c_1z^{d/2}J_{\nu}(k z)+c_2z^{d/2}Y_{\nu}(k z)
\ee 
where we define
\be 
\nu\equiv \frac{\sqrt{d^2+4\mu^2}}{2}
\ee 
and where $k\equiv\sqrt{\abs{k_ik^i}}$.\footnote{We specifically chose the letter $\nu$ to denote $\frac{\sqrt{d^2+4\mu^2}}{2}$ as this term up to an overall $i$ is the pole of the spectral representation of the bulk to bulk propagator, usually denoted as $\pm i\nu$ in the literature \cite{Carmi:2019ocp}, where there are two poles due to the shadow symmetry.}\textsuperscript{,}\footnote{The scaling dimension of the dual operator in the boundary CFT is $\De=\frac{d}{2}+\nu$ in our notation. We would like to caution the reader that many papers in literature calls mass $m$ what we defined as the \emph{effective mass} $\mu$, hence the well-known formula $\De(\De-d)=m^2$, which becomes $\De(\De-d)=\mu^2$ in our notation. As we choose to distinguish mass $m$ and effective mass $\mu$, it is completely consistent in our definition when we say \emph{massless conformally coupled scalar} as $m=0$ despite $\mu^2=\frac{1-d^2}{4}\ne 0$.}

For spacelike momenta, the regularity in the AdS can only be achieved for the particular combination which sums up to the Bessel function of the second kind, i.e.
\be 
\Phi_k(z)\sim z^{d/2}K_\nu(k z)
\ee 
Note that this is always possible due to the identity\footnote{This is only true for $z>0$. For generic $z$, the relevant identity reads as 
\be 
K_{\nu }(z)=\left\{
\begin{aligned}
&i^{\nu } \left((-\log (z)+\log (i z)) J_{\nu }(i z)-\frac{1}{2} \pi  Y_{\nu }(i z)\right)\quad & \text{ for } \nu\in \Z\\
&\frac{1}{2} \pi  \csc (\pi  \nu ) \left(\cos (\pi  \nu ) (i z)^{\nu } z^{-\nu }-(i z)^{-\nu } z^{\nu }\right) J_{\nu }(i z)-\frac{1}{2}
\pi  (i z)^{\nu } z^{-\nu } Y_{\nu }(i z) \quad &\text{ for } \nu\not\in \Z
\end{aligned}\right.
\ee 
}
\be 
K_\nu(z)=\frac{\pi  i^{\nu }}{2}\left(i J_\nu(i z)-Y_\nu(i z)\right)
\ee 

Below, we will focus on massless minimally coupled scalars $\Phi_k^{(m)}$ and massless conformally coupled scalars $\Phi_k^{(c)}$ for which the relevant bulk to boundary propagators read as
\bea[eq: bulk to boundary in AdS]
\Phi_k^{(m)}(z)\sim &z^{d/2}K_{d/2}(k z)\\
\Phi_k^{(c)}(z)\sim &z^{d/2}K_{1/2}(k z)\label{eq: bulk to boundary in AdS for conformally coupled}
\eea 
where we use the fact that Ricci scalar $R=-d(d+1)$ in AdS.

We can similarly calculate the bulk to bulk propagators. We are looking for the solutions to the equation
\be 
\left(z^{1+d}\partial_z z^{1-d}\partial_z-z^{2}k_ik^i-\mu^2\right)G_\Phi(z,z',k_i)=i\delta(z-z')z^{d+1}
\ee 
We observe that
\be
\left(\partial_z z^{1-d}\partial_z-z^{1-d}k_ik^i-z^{-1-d}\mu^2\right)\left(
-\frac{i p z^{d/2} z'^{d/2} J_{\nu }(p z) J_{\nu }(p z')}{k_ik^i+p^2}
\right)=iz^{1-d/2}z'^{d/2}pJ_{\nu }(p z) J_{\nu }(p z')
\ee 
and since we also have the identity
\be 
\int\limits_0^{\infty} pJ_{\nu }(p z) J_{\nu }(p z') dp=\frac{\delta(z-z')}{z}
\ee 
we find the propagator:
\be 
\label{eq: propagator in AdS}
G_\Phi(z,z',k_i)=\int\limits_0^{\infty}\frac{-ipdp}{k_ik^i+p^2-i\epsilon}\left(z^{d/2}J_{\nu }(p z)\right)\left(z'^{d/2}J_{\nu }(p z')\right)
\ee 
In particular, $\nu\rightarrow 1/2\;(d/2)$ gives the propagator for conformally (minimally) coupled scalar as we noted above.

One can now go ahead and write the expression for Witten diagrams. At tree level, the amplitude for a  diagram of $m$ external legs and $n$ bulk propagators reads as
\be 
\label{eq: generic tree level amplitude}
W_{m,n}\sim \int_0^\infty dz_1\dots dz_{n+1}\Phi_{k_1}(z_{i_1})\dots\Phi_{k_m}(z_{i_m})G_\Phi(z_{j_1},z_{j_2},q_1)\dots G_\Phi(z_{j_{n}},z_{j_{n}},q_{n})\prod\limits_{t=1}^{n+1} \rho_{t}(z_t)
\ee 
for the interaction coefficient $\rho_n(z_n)$ at $n^\text{th}$ vertex, where $z_{i_1},\dots,z_{i_m}\in\{z_1,\dots,z_{n+1}\}$ and $z_{j_1},\dots,z_{j_r}\in\{z_1,\dots,z_{n+1}\}$, and where $q_i$ are norms of linear combinations of vectors $\bm{k}_i$ depending on the topology. 

\begin{figure} 
	\centering
	\includegraphics[width=.25\textwidth,origin=c]{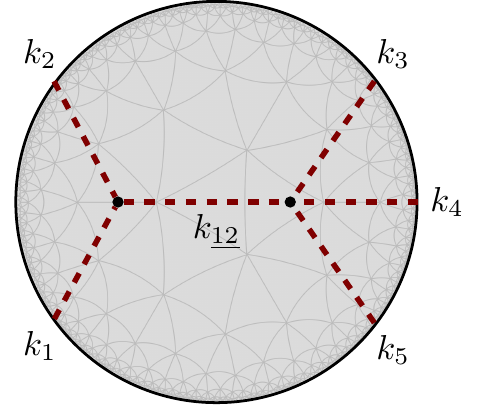}
	\caption{A five point tree level Witten diagram, labeled as $W_{5,1}$ below.\label{fig: 5pt2 for minimal}}
\end{figure}

As an example, we can consider the topology in Figure~\ref{fig: 5pt2 for minimal} for minimally coupled scalars; the relevant amplitude would read as
\be 
W\sim \int_0^\infty dz_1 dz_2\Phi^{(m)}_{k_1}(z_1)\Phi^{(m)}_{k_2}(z_1)\Phi^{(m)}_{k_3}(z_2)\Phi^{(m)}_{k_4}(z_2)\Phi^{(m)}_{k_5}(z_2)G_\phi(z_1,z_2,k_{\underline{12}})\lambda_3\lambda_4
\ee 
where we are following the notation of \cite{Albayrak:2018tam} for addition of $k-$vectors.\footnote{Explicitly:
	\begin{subequations}
		\label{eq: notation}
		\be
		k_{\underline{i_{11}i_{12}\dots i_{1n_1}}\;\underline{i_{21}i_{22}\dots i_{2n_2}}\dots \underline{i_{m1}i_{m2}\dots i_{mn_m}}j_1j_2\dots j_p}\coloneqq\sum\limits_{a=1}^{m}\abs{\sum\limits_{b=1}^{n_a}\bm{k}_{i_{ab}}}+\sum\limits_{c=1}^{p}\abs{\bm{k}_{j_c}}\;,
		\ee
		and
		\be
		\bm{k}_{i_1i_2\dots i_n}\coloneqq\bm{k}_{i_1}+\bm{k}_{i_2}+\cdots+\bm{k}_{i_n}\;.
		\ee
		For example, $k_{\underline{12}3\underline{45}}\equiv\abs{\bm{k}_1+\bm{k}_2}+\abs{\bm{k}_3}+\abs{\bm{k}_4+\bm{k}_5}\;$ and $\bm{k}_{12}\equiv\bm{k}_1+\bm{k}_2$.
	\end{subequations}
}

We can use \equref{eq: propagator in AdS} and exchange the order of integrations, writing \equref{eq: generic tree level amplitude} in a different form:
\be 
\label{eq: prescription}
W_{m,n}\sim \int_0^\infty dp_1\dots dp_{n}\frac{-ip_1}{p_1^2+q_1^2-i\epsilon}\dots \frac{-ip_{n}}{p_{n}^2+q_{n}^2-i\epsilon}\cB_1(k_i,p_i)\dots \cB_{n+1}(k_i,p_i)\prod\limits_{t=1}^{n+1} \rho_{t}
\ee 
where $\cB$ is what we will call bulk-point integrated expression. Note that we integrated (if any) $z-$dependent parts of the interaction coefficients as well.

\subsection{Example: minimally coupled scalars in AdS$_4$}
\label{sec: minimally coupled scalars}
In this section we will calculate some amplitudes for tree level Witten diagrams using the procedure advocated in the previous section; specifically, with
\be 
\phi_k(z)\equiv&\sqrt{\frac{2}{\pi}}(k z)^{3/2}K_{3/2}(k z)\\
G_\phi(z,z',k_i)\equiv&\int\limits_0^{\infty}\frac{-ipdp}{k_ik^i+p^2-i\epsilon}\left(z^{3/2}J_{3/2 }(p z)\right)\left(z'^{3/2}J_{3/2 }(p z')\right)
\ee 
where we choose a particular normalization for the bulk to boundary propagator consistent with the literature.\footnote{\label{footnote about normalization}An overall $k-$dependent scaling of bulk to boundary propagators, i.e. $\phi_k(z)\rightarrow f(k)\phi_k(z)$, is immaterial for our purposes in this paper, hence we refer this as a normalization and fix it with a convenient factor. However, this \emph{normalization} is actually tightly constrained by scaling dimensions of the dual operators at the boundary CFT.} 

Apart from its physical significance, we focus on AdS$_4$ also because Bessel functions simplify for half integer arguments; hence the calculations are relatively easier for AdS$_{2+2n}$. This motivation was also used in previous similar work \cite{Albayrak:2018tam,Albayrak:2019asr,Albayrak:2019yve}, where the calculation of graviton amplitudes in \cite{Albayrak:2019yve} is actually quite similar to the computation at hand. Specially, one can write the graviton propagators in AdS$_{d+1}$ as
\be
h_{ij}(z,k_i)=&\frac{\epsilon_{ij}}{z^2}\phi_k(z)\\
G^{\text{graviton}}_{ab,cd}(z,z',k_i)=&\frac{i}{(z z')^2}\cD^{\bk}_{ab,cd}G_\phi(z,z',k_i)
\ee 
where $\cD^{\bk}_{ab,cd}$ is a differential operator whose details are irrelevant for us. However, one important remark is that these differential operators commute with the rest of the calculation, thus for a Witten diagram with $n$ bulk to bulk propagators we schematically have
\be 
\label{eq: schematic form}
\cA_{\text{graviton}}=\left(\epsilon_i,V_{i}\right)^{a_{11}a_{12}\dots a_{14}a_{21}\dots a_{n4}}\prod\limits_{j=1}^{n}\cD_{a_{j1}a_{j2},a_{j3}a_{j4}}^{\bm{p}_j}\cM\;,
\ee 
where $(\epsilon_i,V_i)$ stand for the collection of the vertex factors and polarization vectors, $\bm{p}_j$ is  sum of some bulk to boundary momenta depending on the topology of the diagram, and $\cM$ is the \emph{scalar factor} of the amplitude. This scalar factor for graviton amplitude is \emph{almost the same expression} with the amplitude for the same Witten diagram with graviton legs replaced by minimally coupled scalars. The only difference between the amplitude for minimally coupled scalars and \emph{graviton scalar factor} is due to the different overall exponent of $z$ in bulk point integation.\footnote{Of course, the vertex coefficients are also different but that is an overall factor which can be easily accounted for.}\textsuperscript{,}\footnote{There are two cases where $z-$factors coincidentally match: minimally coupled scalars with two-derivative-cubic interaction (polynomial quartic interaction) have exactly the same $z-$factor with cubic (quartic) graviton interaction. We will not be making use of that correspondence though, as we are only interested in polynomial scalar interactions and as we do not know of any explicit result in the literature for momentum space Witten diagrams of pure quartic graviton interactions.} 

We now proceed with calculation of bulk point integrated expressions. Specifically, we define
\bea 
\cK\cK\cK(k_1,k_2,k_3)\equiv&\int\limits_0^\infty\frac{dz}{z^4}\phi_{k_1}(z)\phi_{k_2}(z)\phi_{k_3}(z)\\
\cK\cK\cJ(k_1,k_2,p)\equiv&\int\limits_0^\infty\frac{dz}{z^4}\phi_{k_1}(z)\phi_{k_2}(z)\left(z^{3/2}J_{3/2}(p z)\right)\\
\cK\cJ\cJ(k,p_1,p_2)\equiv&\int\limits_0^\infty\frac{dz}{z^4}\phi_{k}(z)\left(z^{3/2}J_{3/2}(p_1 z)\right)\left(z^{3/2}J_{3/2}(p_2 z)\right)\\
\cK\cK\cK\cJ(k_1,k_2,k_3,p)\equiv&\int\limits_0^\infty\frac{dz}{z^4}\phi_{k_1}(z)\phi_{k_2}(z)\phi_{k_3}(z)\left(z^{3/2}J_{3/2}(p z)\right)
\eea 
We can define similar expressions for more complicated interactions, but we will restrict to the first few simplest tree level Witten diagrams. By regularizing the integrations we find
\footnotesize
\begin{subequations}
\begin{align}
\cK\cK\cK(k_1,k_2,k_3)=&\frac{1}{9} \left(k_1+k_2+k_3\right)^3-k_1 k_2 k_3+\frac{1}{3}
\left(k_1^3+k_2^3+k_3^3\right) \left(-\log \left(k_1+k_2+k_3\right)-\gamma +1\right)
\\
\cK\cK\cJ(k_1,k_2,p)=&
\frac{\sqrt{\frac{2}{\pi }} \left(k_1^3+k_2^3\right)
	\left(\tan
	^{-1}\left(\frac{p}{k_1+k_2}\right)-\frac{p}{k_1+k_2}\right
	)}{3 p^{3/2}}-\frac{p^{3/2} \left(3 \log
	\left(\left(k_1+k_2\right){}^2+p^2\right)+6 \gamma
	-8\right)}{9 \sqrt{2 \pi }}
\end{align}
\begin{align}
\cK\cJ\cJ(k,p_1,p_2)=&\frac{k^3 \tanh ^{-1}\left(\frac{2 p_1
		p_2}{k^2+p_1^2+p_2^2}\right)-2 k p_1
	p_2+\left(p_2^3-p_1^3\right) \tan
	^{-1}\left(\frac{p_1-p_2}{k}\right)+\left(p_1^3+p_2^3\right
	) \tan ^{-1}\left(\frac{p_1+p_2}{k}\right)}{3 \pi 
	\left(p_1 p_2\right)^{3/2}}
\\
\cK\cK\cK\cJ(k_1,k_2,k_3,p)=&\frac{\sqrt{\frac{2}{\pi }} \left(k_1^3+k_2^3+k_3^3\right) \tan
	^{-1}\left(\frac{p}{k_1+k_2+k_3}\right)}{3 p^{3/2}}-\frac{p^{3/2} \left(3 \log
	\left(\left(k_1+k_2+k_3\right){}^2+p^2\right)+6 \gamma -8\right)}{9 \sqrt{2 \pi }}\nn\\&-\sqrt{\frac{2}{\pi p}}
\left(\frac{k_1 k_2 k_3
	\left(k_1+k_2+k_3\right)}{\left(k_1+k_2+k_3\right){}^2+p^2}+\frac{k_1^3+k_2^3+k_3^3-3 k_1k_2 k_3}{3 \left(k_1+k_2+k_3\right)}\right)
\end{align}
\end{subequations}
\normalsize
where $\gamma$ is the Euler-gamma number.

\begin{figure} 
	\centering
	$\begin{aligned}
	\includegraphics[width=.15\textwidth,origin=c]{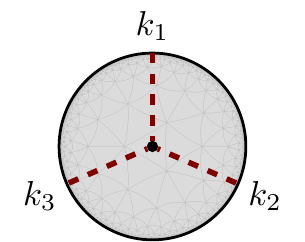}
	\end{aligned}$
	$\begin{aligned}
	\includegraphics[width=.20\textwidth,origin=c]{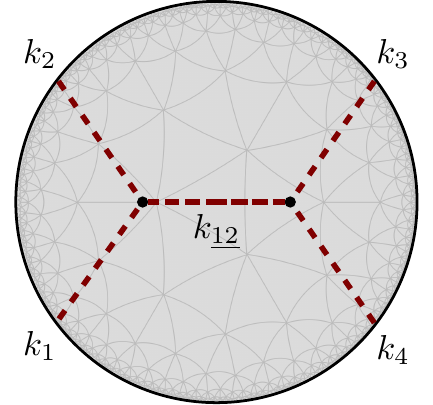}
	\end{aligned}$
	$\begin{aligned}
	\includegraphics[width=.20\textwidth,origin=c]{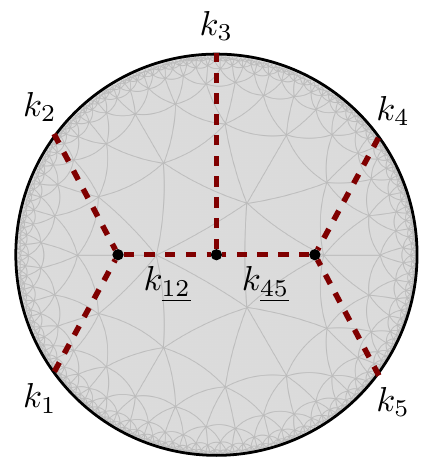}
	\end{aligned}$
	$\begin{aligned}
	\includegraphics[width=.25\textwidth,origin=c]{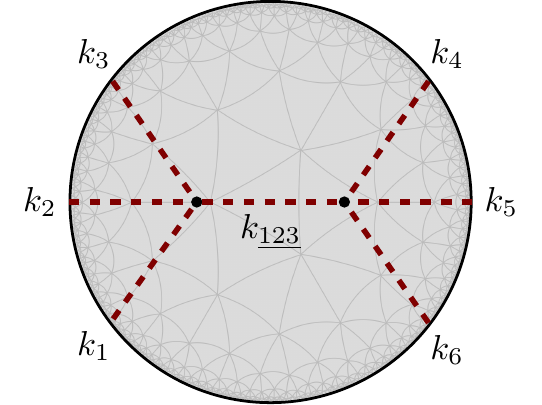}
	\end{aligned}$
	$\begin{aligned}
\includegraphics[width=.24\textwidth,origin=c]{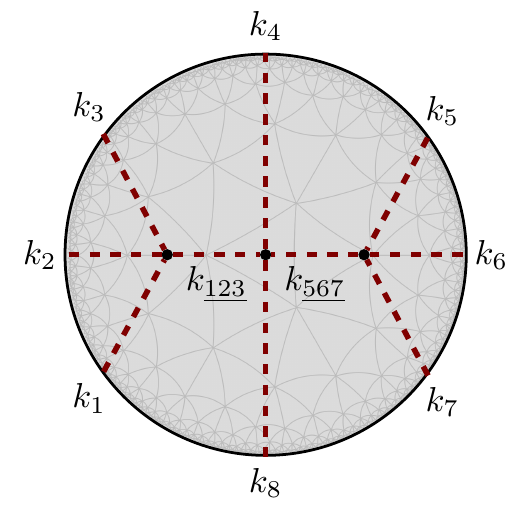}
\end{aligned}$
	$\begin{aligned}
\includegraphics[width=.24\textwidth,origin=c]{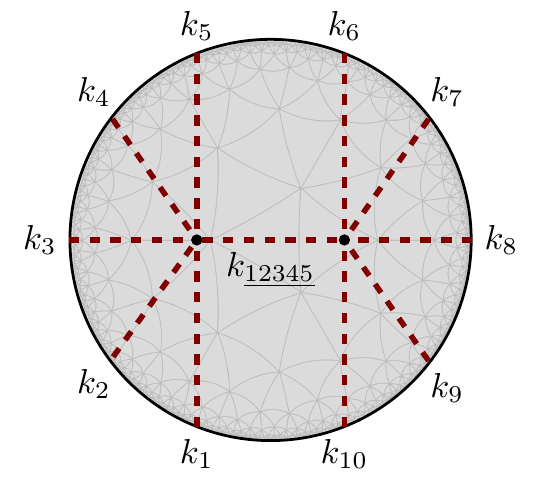}
\end{aligned}$
	$\begin{aligned}
\includegraphics[width=.24\textwidth,origin=c]{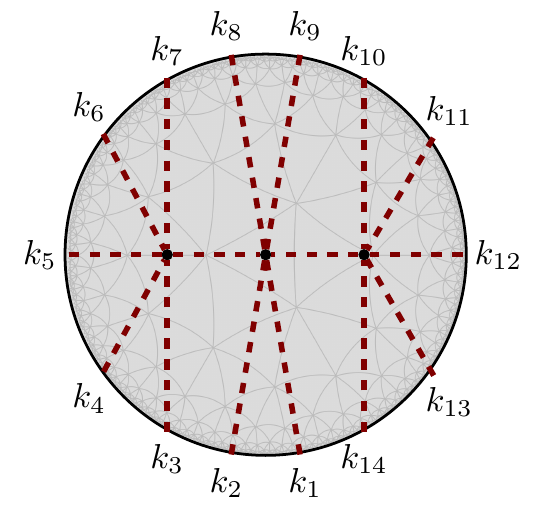}
\end{aligned}$
	$\begin{aligned}
\includegraphics[width=.24\textwidth,origin=c]{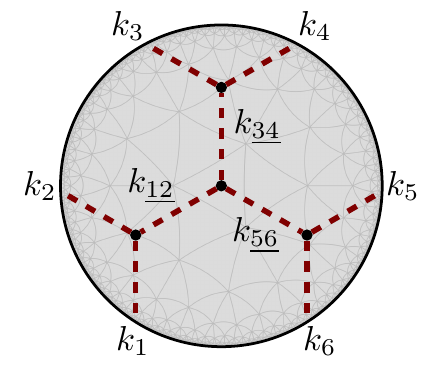}
\end{aligned}$
	\caption{Various tree level Witten diagrams that will be of interest below. From left to right, we label them as $W_{3,0}$, $W_{4,1}$, $W_{5,2}$, $W_{6,1}$, $W_{8,2}$, $W_{10,1}$, $W_{14,2}$, and $W_{6,3}$.\label{fig: minimally coupled scalar}}
\end{figure}

We can now use the prescription of \equref{eq: prescription} to write down the amplitudes for various Witten diagrams:
\bea[eq: witten amplitudes for minimally coupled scalars]
W_{3,0}(k_i)=&-i\lambda_3  \cK\cK\cK(k_1,k_2,k_3)\\
W_{4,1}(k_i)=&-\lambda_3^2\int_0^\infty dp\frac{-ip}{p^2+k_{\underline{12}}^2-i\epsilon}\cK\cK\cJ(k_1,k_2,p)\cK\cK\cJ(k_3,k_4,p)\\
W_{5,1}(k_i)=&-\lambda_3\lambda_4\int_0^\infty dp\frac{-ip}{p^2+k_{\underline{12}}^2-i\epsilon}\cK\cK\cJ(k_1,k_2,p)\cK\cK\cK\cJ(k_3,k_4,k_5,p)\\
W_{5,2}(k_i)=&i\lambda_3^3\int_0^\infty dp_1dp_2\frac{-ip_1}{p_1^2+k_{\underline{12}}^2-i\epsilon}\frac{-ip_2}{p_2^2+k_{\underline{45}}^2-i\epsilon}\nn\\&\qquad\qquad\qquad\x\cK\cK\cJ(k_1,k_2,p_1)\cK\cJ\cJ(k_3,p_1,p_2)\cK\cK\cJ(k_4,k_5,p_2)\\
W_{6,1}(k_i)=&-\lambda_4^2\int_0^\infty dp\frac{-ip}{p^2+k_{\underline{123}}^2-i\epsilon}\cK\cK\cK\cJ(k_1,k_2,k_3,p)\cK\cK\cK\cJ(k_4,k_5,k_6,p)
\eea 

Clearly, these are hard, albeit doable, integrals.\footnote{One of the key points of \cite{Albayrak:2018tam} where the authors computed similar integrals for gluon exchange is that one can use residue theorem to significantly simplify such formidable integrals. Unfortunately, the integrands in \equref{eq: witten amplitudes for minimally coupled scalars} do not fall off at infinity hence residue theorem is no longer a simple option.}  However, we will not dwell on these integrals for two reasons: the first reason is the simplicity of the computation of conformally coupled scalars compared to that of minimally coupled scalars. This is a fortunate observation because conformally coupled scalars can potentially be used as seed diagrams from which minimally coupled scalar can be computed as well \cite{Baumann:2019oyu}.\footnote{
In addition to those motivations, we also would like to note the claim of \cite{Faraoni:2013igs} that massive scalars in curved background may lie on the lightcone in the local Minkowski frame unless they are conformally coupled, leading to causal pathologies, indicating that any massive scalar needs to be conformally coupled.} And that is what we are turning to in next section.

\section{Conformally coupled scalars}
\label{sec: conformally coupled scalars}
\subsection{An algorithmic approach for conformally invariant scalars}
\label{sec:algorithm}
In \secref{\ref{sec: scalars in curved spacetime}} we used a Weyl transformation to deduce which interaction terms preserve conformal invariance by going to the flat space and checking the form of interaction coefficient. However, we can use that Weyl transformation for computation purposes as well. Indeed, for conformally coupled scalars\footnote{In the rest of the paper we mean $\xi=\xi_c$ and $m=0$ when we say conformally coupled scalar.}, the Lagrangian in \equref{eq: polynomial lagrangian} simplifies to
\be 
S=-\int d^dx dz \left[\half\left(\partial_i\phi\right)^2+\half\left(\partial_z\phi\right)^2+\frac{\lambda_n(z)}{n!} \phi^n\right]
\ee 
under the transformation in \equref{eq:Weyl transformation}. Here, we defined
\be 
\label{eq: definition of lambda z}
\lambda_n(z)\equiv \lambda_nz^{\frac{d-1}{2}\left(n-n_c\right)}
\ee 
where $i=1,\dots,d$ run for the boundary coordinates with the boundary metric $\eta_{ij}$. From eqns.~(\ref{eq: field weyl transformation}, \ref{eq: bulk to boundary in AdS for conformally coupled}, \ref{eq: propagator in AdS}), we can immediately write down the propagators:\footnote{As we explained in footnote~\ref{footnote about normalization}, we treat the overall $k-$dependence as normalization which we chose for the conformally coupled scalars in a consistent fashion with the similar work in \cite{Raju:2012zs,Raju:2012zr,Albayrak:2019asr}.}
\bea 
\phi_k(z)= &\sqrt{\frac{2kz}{\pi}}K_{1/2}(k z)\label{eq: bulk to boundary propagator in flat space}\\
G_\phi(z,z',k_i)=&\int\limits_0^{\infty}\frac{-ipdp}{k_ik^i+p^2-i\epsilon}\left(z^{1/2}J_{1/2 }(p z)\right)\left(z'^{1/2}J_{1/2 }(p z')\right)
\eea
from which we can deduce the bulk point integrations:
\be 
\label{eq: bulk point expression for conformally coupled scalars}
\cB_{n,x}(k,p)\equiv\int\limits_0^{\infty}dz z^{\frac{d-1}{2}\left(n-n_c\right)}\prod\limits_{i=1}^x\prod\limits_{j=1}^{n-x}\phi_{k_i}(z)\left(z^{\half}J_{\half}(p_j z)\right)
\ee 

Carrying out such integrals once and for all and then using those results in various different Witten diagrams is part of the strategies that were employed in \cite{Albayrak:2018tam, Albayrak:2019yve} as we demonstrated in the case of minimally coupled scalars in \secref{\ref{sec: minimally coupled scalars}}. However, one can do better than calculating these integrals generically and using them case by case: we can directly find an algebraic algorithm and bypass all integrations, both the bulk-point $z-$integrations and propagator $p-$integrations!

Such an algorithm is discussed in \cite{Albayrak:2019asr} where the authors refer to the \emph{additive} property of vertices, enabling them to work at the level of truncated diagrams and compute amplitudes directly via algebraic means. This is possible, as they argue, because the gluon propagators in AdS$_4$ are simply exponentials and there is a nice cancellation between the volume factor $z^{-d-1}$ and the vertex factor $z^4$ in AdS$_4$. We see that the propagators of conformally coupled scalars in flat space precisely match gluons in AdS$_4$ and bulk point expressions have exactly same $z-$powers if $n=n_c$ as can be seen from \equref{eq: bulk point expression for conformally coupled scalars}. So we arrive at the conclusion that one can reduce the integrations to algebraic calculations for scalars with conformal symmetry in any dimension, analogous to gluons in AdS$_4$.\footnote{One can in fact still use the algorithm for conformally coupled scalars with $n\ne n_c$ with appropriate modification. We will discuss this in next section.}\textsuperscript{,}\footnote{This result is hardly surprising as the algorithm used in \cite{Albayrak:2019asr} is in fact derived by Arkani-Hamed et al in \cite{Arkani-Hamed:2017fdk} for conformally coupled scalars in dS. However the authors actually use the modified version of the algorithm that we will see in \secref{\ref{sec: generalization of algorithm}}, hence they are not really trading all relevant integrations by an algebraic calculation. On the contrary, we will get rid of all integrals in this section, analogous to the case of gluons in AdS$_4$.}

Let us quickly review the algorithm to compute the Witten diagram amplitude of the form
\be 
W_{m,n}\equiv\int_0^\infty dp_1\dots dp_{n}\frac{-ip_1}{p_1^2+q_1^2-i\epsilon}\dots \frac{-ip_{n}}{p_{n}^2+q_{n}^2-i\epsilon}\cB_{n_c,x_1}(k_i,p_i)\dots \cB_{n_c,x_{n+1}}(k_i,p_i)\prod\limits_{t=1}^{n+1} (-i\rho_t)
\ee
which is the expression for the diagram of $m$ external legs $n$ bulk propagators (dependence on external legs is implicit in $\cB$).

\begin{figure}
	\centering
	\includegraphics[width=.7\textwidth,origin=c]{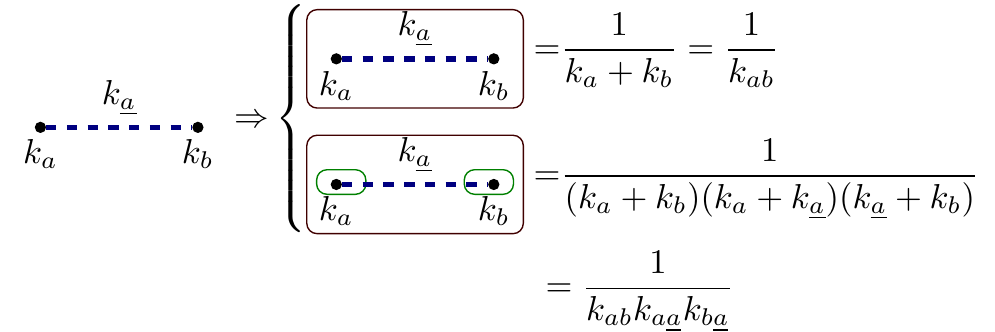} 
	\includegraphics[width=.9\textwidth,origin=c]{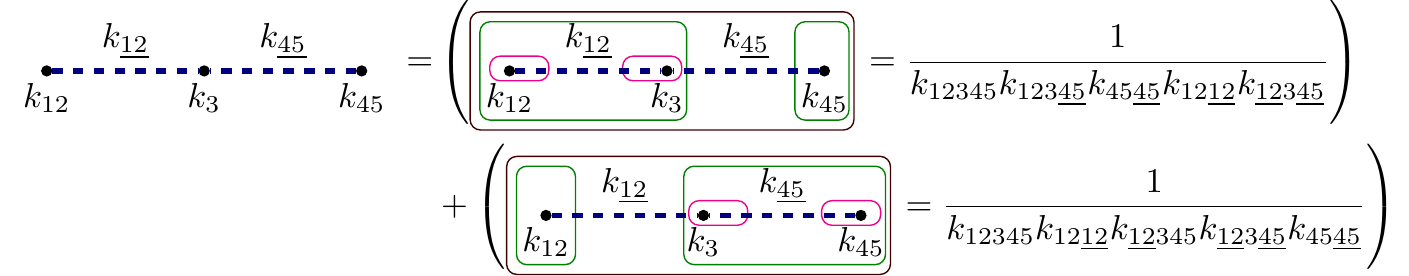}
	\caption{\label{fig: algorithm}Diagrammatic illustration of the algorithm}
\end{figure}

We note that the aforementioned additive property of the vertices, which follows from $\phi_{k_1}(z)\phi_{k_2}(z)=\phi_{k_1+k_2}(z)$, means that we can change the number of external legs as we wish as long as the sum of norms of the momenta flowing to vertices stay the same, up to the change in the coupling coefficients. Hence, we will work with the truncated diagram of the amplitude
\be 
\label{eq: truncated amplitude vs Witten diagram}
\cA_{n}\equiv \frac{i^{2n+1}W_{m,n}}{\prod\limits_{t=1}^n \rho_{t}}
\ee 
which only depends on the topology of the truncated diagram, independent of the details of external legs but only the sum of norms of the incoming momenta.\footnote{We introduced an additional $i^{n}$ factor for convenience; this way, our truncated amplitudes are \emph{exactly same} with those of \cite{Albayrak:2019asr} in which $i$ factors are included in the \emph{projectors} $\Pi$ instead of the scalar part of the propagator.} For example, for\footnote{In \figref{\ref{fig: algorithm}} and \figref{\ref{eq: loop algorithm}}, we use the diagrammatic notation for truncated amplitudes in the same sense they are used in \cite{Albayrak:2019asr}: they correspond to Witten diagrams with bulk to boundary propagators stripped off.} $\cA_1(k_a,k_b,k_{\underline{a}})$, which is given in the first diagram of \figref{\ref{fig: algorithm}}, the amplitudes of the four and six point diagrams shown in \figref{\ref{fig: minimally coupled scalar}} for conformally coupled scalars read as 
\be 
\label{eq: witten41 and witten61}
W_{4,1}(k_i)=&i\lambda_3^2 \cA_1(k_{12},k_{34},k_{\underline{12}})\\
W_{6,1}(k_i)=&i\lambda_4^2 \cA_1(k_{123},k_{456},k_{\underline{123}})
\ee 

The algorithm for the computation of $\cA$ is as follows. The diagram is decomposed into sub-diagrams by \emph{cutting} all internal lines. One then considers all possible orders in which the lines are cut, and assigns partial amplitudes to individual cases. The sum of these partial amplitudes give the full amplitude of the initial diagram.

The partial amplitude for a diagram with its lines cut in a particular order is the product of the amplitudes for all subgraphs, which are in turn equal to the inverse of the sum of all \emph{vertex norms} within that subgraph and \emph{line norms} going out of that subgraph.  

In \figref{\ref{fig: algorithm}}  we illustrate the algorithm for $\cA_1$ and $\cA_2$. For $\cA_1$, we observe that there is only one partial amplitude which yields remarkably simple results
\be 
W_{4,1}(k_i)=\frac{i\lambda_3^2}{k_{1234}k_{12\underline{12}}k_{34\underline{12}}}
\;,\qquad
W_{6,1}(k_i)=\frac{i\lambda_4^2}{k_{123456}k_{123\underline{123}}k_{456\underline{123}}}
\ee 
For $\cA_2$, we have two partial amplitudes where the sum simplifies quite nicely, yielding
\be 
\label{eq: witten52}
W_{5,2}(k_i)=-i\lambda_3^3 \cA_2\left(k_{12},k_{3},k_{45},k_{\underline{12}},k_{\underline{45}}\right)=-i\lambda_3^3\frac{k_{12\underline{12}3345\underline{45}}}{k_{12345}k_{12\underline{12}}k_{345\underline{12}}k_{\underline{12}3\underline{45}}k_{45\underline{45}}k_{123\underline{45}}}
\ee 

We stated above that this algorithm is valid if $n=n_c$. By imposing $n\in\Z$ in \equref{eq: n_c}, we see that there are only three cases with conformally invariant interactions: AdS$_{3,4,6}$ with $\phi^{6,4,3}$ interaction. As the algorithm we provided is independent of the spacetime dimension, we can use it for all truncated diagrams; below we list some results for various Witten diagrams: one should understand the relevant dimension for which the amplitude is valid from the form of the interaction, i.e. results with $4-$point interactions are valid for AdS$_4$ only.\footnote{We provide  the results without dwelling on the relevance of the specific models. In particular, one can see $\phi^3$ potential in AdS$_6$ as a mere toy model due to the $\Z_2$ odd potential yielding a Hamiltonian unbounded from below.}
\bea 
W_{10,1}(k_i)=&i\lambda_6^2\; \cA_1\left(k_{12345},k_{6789(10)},k_{\underline{12345}}\right)
\\
W_{8,2}(k_i)=&-i\lambda_4^3\;
\cA_2\left(k_{123},k_{48},k_{567},k_{\underline{123}},k_{\underline{567}}\right)
\\
W_{14,2}(k_i)=&-i\lambda_6^3\;
\cA_2\left(k_{34567},k_{1289},k_{(10)(11)(12)(13)(14)},k_{\underline{34567}},k_{\underline{(10)(11)(12)(13)(14)}}\right)
\eea 
where
\bea 
\cA_1(q_1,q_2,q_3)= & \frac{1}{(q_1+q_2)(q_1+q_3)(q_2+q_3)}\\
\cA_2(q_1,q_2,q_3,q_4,q_5)= & 
\frac{(q_1+2q_2+q_3+q_4+q_5)}{(q_1+q_2+q_3)(q_1+q_4)(q_2+q_3+q_4)(q_2+q_4+q_5)(q_3+q_5)(q_1+q_2+q_5)}
\eea
We remind the reader of our notation given in \equref{eq: notation}. For example, $k_{(10)(11)(12)(13)(14)}$ above stands for $k_{10}+k_{11}+k_{12}+k_{13}+k_{14}$.

We would like to note that the method is not restricted to comb-like diagrams, and can be used for other topologies as well. For example, for star diagram in  \figref{\ref{fig: minimally coupled scalar}}, the algorithm yields
\be 
W_{6,3}=\Big(\mathcal{I}\,+\, 34\leftrightarrow 56\Big)+
\left(\scriptsize\begin{aligned}
	12\rightarrow 34\\
	34\rightarrow 56\\
	56\rightarrow 12
\end{aligned}\normalsize\right)
+\left(\scriptsize\begin{aligned}
	12\rightarrow 56\\
	56\rightarrow 34\\
	34\rightarrow 12
\end{aligned}\normalsize\right)
\ee 
where
\be 
\mathcal{I}=\frac{i\lambda_3^3}{k_{123456}k_{12\underline{12}}k_{\underline{12}3456}k_{\underline{12}\,\underline{34}\,\underline{56}}k_{34\underline{34}}k_{\underline{12}56\underline{34}}k_{56\underline{56}}}
\ee 
whose  step by step computation can be found in \cite{Albayrak:2019asr}.

\subsection{Generalized algorithm for all conformally coupled scalars}
\label{sec: generalization of algorithm}
There are not so many theories of interacting scalars with full conformal symmetry; in fact, the \equref{eq: n_c} tightly constraints the possibilities into three cases: AdS$_3$ with $\phi^6$, AdS$_4$ with $\phi^4$, and AdS$_6$ with $\phi^3$ as we stated in previous section. However, we can extend our algorithm to all conformally coupled  scalars which are not necessarily invariant under conformal transformations. 

The restriction to conformally invariant scalars followed from the requirement to get rid of the additional $z-$factors in the bulk point integration in \equref{eq: bulk point expression for conformally coupled scalars}: we could use the algorithm if we were to expand additional $z-$factors in terms of $\phi_k(z)$. From \equref{eq: bulk to boundary propagator in flat space}, we observe that this is indeed possible if we expand the interaction coefficients via Laplace transform, i.e.\footnote{Here $\gamma$ is an arbitrary positive constant chosen so that the contour of integration lies to the right of all singularities in $\lambda_n(z)$.}
\be 
\tl\lambda_n(\omega)\equiv\int\limits_{\gamma-i\infty}^{\gamma+i\infty}e^{\omega z}\lambda_n(z)\frac{dz}{2\pi i}\;,\quad \lambda_n(z)=\int\limits_{0}^{\infty}e^{- \omega z}\tl\lambda_n(\omega)d\omega= \int\limits_{0}^{\infty}\phi_{\omega}(z)\tl\lambda_n(\omega)d\omega
\ee 

Thus, we can rewrite \equref{eq: generic tree level amplitude} as
\be 
\label{eq: witten amplitude in terms of laplace transform}
W_{m,n}= \int\limits_{0}^{\infty}d\omega_1\dots d\omega_{n+1}\prod\limits_{t=1}^{n+1}\tl\rho_t(\omega_t) \tl W_{m,n}
\ee 
where $\tl\rho_m$ is the appropriate $\tl\lambda_n(\omega_m)$ at $m^\text{th}$ vertex. Here, we defined
\begin{multline}
\tl W_{m,n}\equiv  \int_0^\infty dz_1\dots dz_{n+1}G_\phi(z_{j_1},z_{j_2},q_1)\dots G_\phi(z_{j_n},z_{j_n},q_n)\\\x
\phi_{k_1}(z_{i_1})\dots\phi_{k_m}(z_{i_m})\phi_{\omega_1}(z_1)\dots \phi_{\omega_{n+1}}(z_{n+1})
\end{multline}
which exactly has the required form hence can be computed by mere algebraic means as we reviewed in the previous section.

Clearly, this modified algorithm is not as efficient as the original one because we still have to compute integrals to get the tree-level AdS amplitudes. However, for a Witten diagram of $n$ vertices, we are replacing $2n-1$ integrations\footnote{$n-1$ $p-$integrations for the bulk to bulk propagators and $n$ $z-$integrations as bulk-point integrations.} with $n$ integrations whose integrand is computed algebraically; so this approach becomes rewarding especially as we consider higher order amplitudes.\footnote{One might object that this naive counting of integrals is misleading as we also need to compute $\tl\lambda_n(\omega)$, the inverse Laplace transform of $\lambda_n(z)$. However, $\lambda_n(z)$ has a pure power law dependence for both polynomial and derivative interactions, hence its inverse Laplace transform is quite trivial,  i.e. $\tl\lambda_n(\omega)=\frac{\omega^{-1-k}}{\G(-k)}$ for $\lambda_n(z)=z^k$.}

Apart from introducing a uniform treatment for all conformally coupled scalars, generalizing the algorithm as above can reveal algebraic and recursive relations between various Witten diagrams. In fact, this way of rewriting an amplitude is already done in \cite{Arkani-Hamed:2017fdk} where they write the cosmological wavefunction $\tl \psi$ for conformally coupled scalar as an integral over the modified wavefunction $\psi$ which follows from the Fourier expansion of the coupling coefficient $\lambda$.\footnote{Our $W$ and $\tl W$ are analogous to their $\tl \psi$ and $\psi$ respectively. Likewise, their equation 2.9 is the analog of our 2.38.} For the utility of such a representation in terms of algebraic \& recursive means and relations with polytopes, we refer the reader to their paper.

As an example, we can consider $\phi^3$ interaction in AdS$_4$. We can immediately read off $\tl W$ from
\equref{eq: witten41 and witten61} and \equref{eq: witten52} by including additional $\omega_j$ dependencies:
\be 
\tl W_{4,1}(k_i,\omega_j)=&i\lambda_3^2 \cA_1(k_{12}+\omega_1,k_{34}+\omega_2,k_{\underline{12}})
\\
\tl W_{5,2}(k_i,\omega_j)=&-i\lambda_3^3 \cA_2\left(k_{12}+\omega_1,k_{3}+\omega_2,k_{45}+\omega_3,k_{\underline{12}},k_{\underline{45}}\right)
\ee 
The amplitude for the relevant Witten diagrams become
\be 
W_{4,1}=\int\limits_0^\infty d\omega_1 d\omega_2\tl W_{4,1}\;,\quad 
W_{5,2}=\int\limits_0^\infty d\omega_1 d\omega_2d\omega_3\tl W_{5,2}
\ee 
Here we used the fact that the interaction coefficient is $\frac{-i\lambda_3}{z}$
whose numerator is taken into account in the calculation of $\tl W$, hence 
\be 
\lambda_3(z)=z^{-1}\quad\Rightarrow\quad\tl\lambda_3(\omega)=1
\ee 

Utilizing softwares for symbolic computations, such as \texttt{Mathematica}, we can calculate such integrals relatively easily. For example, $W_{4,1}$ reads as
\be 
W_{4,1}=-\frac{i \lambda ^2}{4 k_{\underline{12}}}&\bigg(
2 \text{Li}_2\left(\frac{k_{1234}}{k_{12}-k_{\underline{12}}}\right)-2 \text{Li}_2\left(\frac{k_{1234}}{k_{12\underline{12}}}\right)+\log
^2\left(\frac{1}{k_{\underline{12}}-k_{12}}\right)+\log ^2\left(k_{12\underline{12}}\right)\\&+2 \log \left(\frac{k_{\underline{12}}-k_{34}}{k_{12\
		\underline{12}}}\right) \log \left(\frac{k_{12\underline{12}}}{k_{1234}}\right)-2 \log \left(k_{12\underline{12}}\right) \log \left(k_{34	\underline{12}}\right)\\&+2 \log \left(k_{1234}\right) \log \left(\frac{k_{34\underline{12}}}{k_{\underline{12}}-k_{12}}\right)+\pi ^2
\bigg)
\ee 
One can compute other tree level expressions $W_{m,n}$ in a similar fashion.

\subsection{Extension to loops}

\begin{figure} 
	\centering
	\includegraphics[width=.25\textwidth,origin=c]{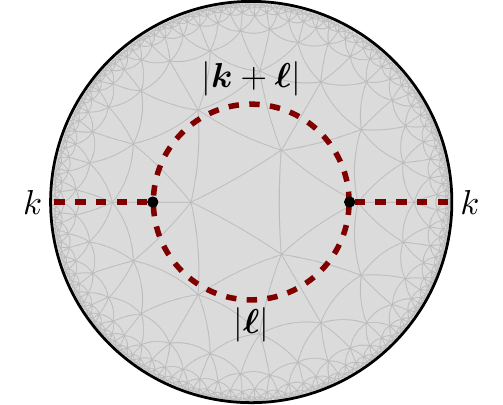}
	$\qquad\quad$
	\includegraphics[width=.25\textwidth,origin=c]{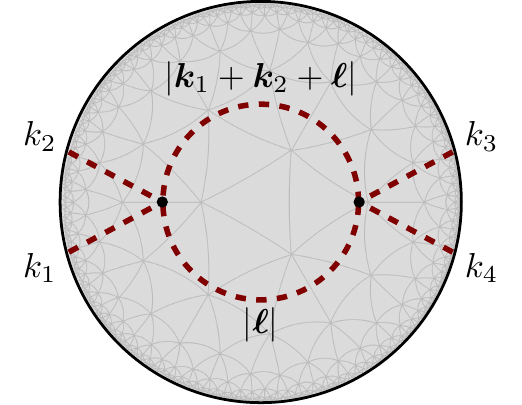}
	\caption{One loop correction to the propagator and four point interaction in AdS\label{fig: loops}}
\end{figure}

The algorithm we presented in \secref{\ref{sec:algorithm}}, and its extension in \secref{\ref{sec: generalization of algorithm}} readily applies to loops as well. In \cite{Arkani-Hamed:2017fdk}, such calculations are already done in the context of cosmological wavefunctions; here, we will illustrate the algorithm in the simplest case of one loop with two vertices and use it to calculate 1 loop corrections to scalar amplitudes in AdS.\footnote{We refer the reader to \cite{Giombi:2017hpr} for analysis of similar loop diagrams of conformally coupled scalars dual to vacuum correlator instead of transition amplitudes.}

\begin{figure}
	\centering
	\includegraphics[width=.9\textwidth,origin=c]{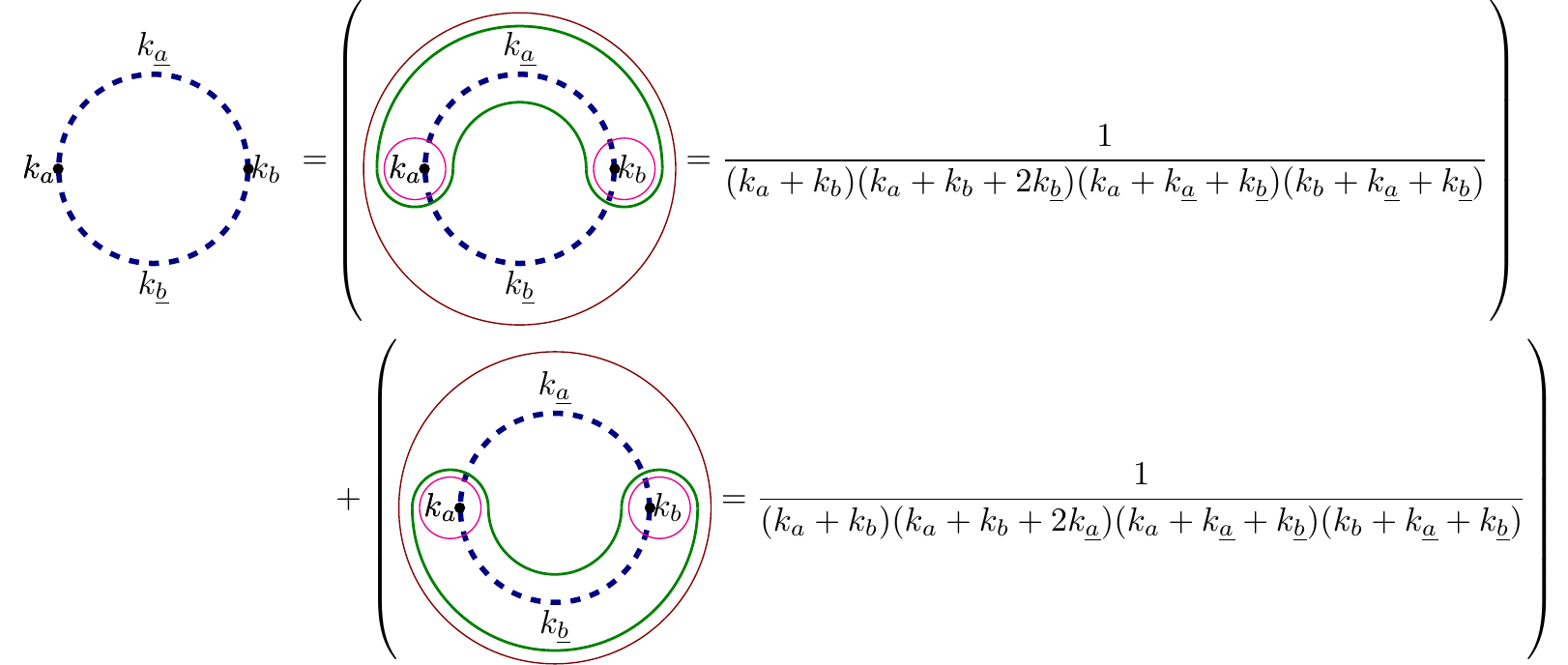}
	\caption{\label{eq: loop algorithm} The application of the algorithm to one loop two vertex diagram}
\end{figure}

The application of the algorithm in case of loops is same with that of tree level diagrams: we decompose the amplitude into sum of partial amplitudes where the expression for each partial amplitude follows from the vertex momenta inside the diagrams and the propagators going out of the diagrams. As we can see  in \figref{\ref{eq: loop algorithm}}, there are two partial amplitudes for one loop two vertex diagram. We multiply the corresponding expressions for each subdiagram in the denominator; we listed them from outward to inward: the first value, $(k_a+k_b)$, corresponds to the big circle, and the last ones correspond to small pink circles. In the end, the final expression takes a rather compact form in our notation: $\left(k_{ab}k_{\underline{a}a\underline{b}}k_{\underline{a}b\underline{b}}\right)^{-1}\left(k_{\underline{a}ab\underline{a}}^{-1}+k_{\underline{b}ab\underline{b}}^{-1}\right)$.

If we were dealing with a tree-level expression, we were done: we could simply translate from this truncated amplitude to a Witten diagram via  \equref{eq: truncated amplitude vs Witten diagram}. However, $k_{\underline{a}}$ and $k_{\underline{b}}$ depends on the loop amplitude and one needs to integrate it as well.

Let us illustrate this in case of loop correction to a two point function, the first diagram in \figref{\ref{fig: loops}}, where we are assuming $\phi^3$ interaction in AdS$_{6}$.\footnote{As we saw in \secref{}, $d=5$ is the only case for which  $\phi^3$ interaction can be calculated without additional $\omega$-integrations, which we would like to avoid to give the simplest example as the illustration.} In this simple case the integrand becomes
\be 
\text{Integrand}=\frac{1}{2k(k+\abs{\bm{l}}+\abs{\bm{k}+\bm{l}})^2}\left(\frac{1}{2k+2\abs{\bm{l}}}+\frac{1}{2k+2\abs{\bm{k}+\bm{l}}}\right)
\ee 
which is to be integrated over $\bm{l}\in \R^5$. As the integration domain is invariant under successive applications of $\bm{l}\rightarrow-\bm{l}$ and $\bm{l}\rightarrow\bm{l+k}$, which interchanges first and second term above, we can write the truncated amplitude as 
\be 
\cA^L(k,k,\bm{k})=\int d^5\bm{l} \frac{1}{2k(k+\abs{\bm{l}}+\abs{\bm{k}+\bm{l}})^2(k+\abs{\bm{l}})}
\ee 
which reads in spherical coordinates after the Wick rotation as
\be 
\cA^L(k,k,\bm{k})=-\frac{i S_3}{2k}\int\limits_0^{\pi}\sin\theta d\theta\int\limits_0^\infty l^4dl\frac{1}{\left(k+l+\sqrt{k^2+l^2+2kl\cos\theta}\right)^2(k+l)}
\ee  
where $S_n=\frac{2\pi^{\frac{n+1}{2}}}{\G\left(\frac{n+1}{2}\right)}$ is the volume of $n-$sphere. We can carry out the $\theta-$integration immediately, yielding
\be 
\cA^L(k,k,\bm{k})=-\frac{i S_3}{2k}\int\limits_0^\infty dl \frac{l^3 \left(-\frac{k+l}{\left| k-l\right| +k+l}+\log \left(\frac{2 (k+l)}{\left| k-l\right| +k+l}\right)+\frac{1}{2}\right)}{k (k+l)}
\ee  
Clearly, this is a divergent integral. By regularizing it with a hard cut-off $\Lambda$\footnote{We thank Aaron Hillman for pointing out that one may need to choose $z-$dependent hard cut-off $\Lambda$ as energy scales vary with the bulk radius, and we believe that choosing $\Lambda/z$ should yield a more uniform energy cut-off. Nevertheless, our calculation should be fine for the purpose of illustrating the usage of algorithm with the loops.}, we find 
\be 
\cA^L(k,k,\bm{k})=-\frac{i\pi ^2 }{72} \left(\frac{18 \Lambda ^2}{k}+96 k \log \left(\frac{\Lambda }{k}\right)+\left(39-6 \pi ^2\right) k-72 \Lambda \right)
\ee 

In the general case of $\phi^{n+2}$ interaction in AdS$_{d+1}$, we have 
\be 
k_a=&\abs{\bm{k}_1}+\dots+\abs{\bm{k}_n}+\omega_1\\
k_b=&\abs{\bm{k}_{n+1}}+\dots+\abs{\bm{k}_{2n}}+\omega_2\\
k_{\underline{a}}=&\abs{\bm{l}}\\
k_{\underline{b}}=&\abs{\bm{l}+\bm{k}_1+\dots+\bm{k}_n}\\
\ee 
which means $k_{\underline{a}ab\underline{a}}\leftrightarrow k_{\underline{b}ab\underline{b}}$ under the successive applications of $\bm{l}\rightarrow -\bm{l}$ and \mbox{$\bm{l}\rightarrow\bm{l}+\bm{k}_1+\dots+\bm{k}_n$} under which the integration is invariant. Therefore, we have
\begin{multline}
\cA^L\left(k_a,k_b,\bm{k}_1+\dots+\bm{k}_n\right)= -\frac{2i S_{d-2}}{k_{12\dots {(2n)}}+\omega_1+\omega_2}\int\limits_0^{\pi}\sin\theta d\theta\int\limits_0^\infty l^{d-1}dl\frac{1}{k_{12\dots {(2n)}}+\omega_1+\omega_2+2l}
\\\x\frac{1}{k_{12\dots n}+\omega_1+l+\sqrt{l^2+k_{\underline{12\dots n}}^2+2 l k_{\underline{12\dots n}}\cos \theta }}
\\\x\frac{1}{k_{(n+1)(n+2)\dots (2n)}+\omega_2+l+\sqrt{l^2+k_{\underline{12\dots n}}^2+2 l k_{\underline{12\dots n}}\cos \theta }}
\end{multline}
whose $\theta$ integration can be immediately carried out:
\begin{multline}
\cA^L\left(k_a,k_b,\bm{k}_1+\dots+\bm{k}_n\right)=- \frac{4 i \pi ^{\frac{d-1}{2}}}{\Gamma \left(\frac{d-1}{2}\right)}\\\x\int\limits_0^\infty dl 
\frac{ l^{d-2}\left(\left(k_a+l\right) \log \left(\frac{k_a+2 l+k_{\underline{12\dots n}}}{k_a+\left| l-k_{\underline{12\dots n}}\right| +l}\right)-\left(k_b+l\right) \log \left(\frac{k_b+2 l+k_{\underline{12\dots n}}}{\left| l-k_{\underline{12\dots n}}\right|
		+k_b+l}\right)\right)}{k_{\underline{12\dots n}} \left(k_a-k_b\right) \left(k_a+k_b\right) \left(k_a+k_b+2 l\right)}
\end{multline}
Despite the term $k_a-k_b$ in the denominator, the integrand above is actually continuous at $k_a=k_b$.

We were not able to compute the integration above for generic $d$, however, it is quite doable once we restrict to a specific $d$; for example,
\begin{multline}
\cA^L\left(k_a,k_b,\bm{k}_1+\dots+\bm{k}_n\right)\evaluated_{d=3} = 
-\frac{i\pi}{k_a+k_b}\left[\log(2)+\frac{k_b \log \left(\frac{k_a+k_{\underline{12\dots n}}}{\Lambda }\right)-k_a \log \left(\frac{k_b+k_{\underline{12\dots n}}}{\Lambda }\right)}{k_a-k_b}\right]
\\
-\frac{i\pi}{2 k_{\underline{12\dots n}}}\Bigg[
\text{Li}_2\left(-\frac{k_a+k_{\underline{12\dots n}}}{k_b-k_{\underline{12\dots n}}}\right)+\text{Li}_2\left(-\frac{k_b+k_{\underline{12\dots n}}}{k_a-k_{\underline{12\dots n}}}\right)+\log \left(k_a+k_{\underline{12\dots n}}\right) \log \left(k_b+k_{\underline{12\dots n}}\right)
\\
-\log \left(k_a+k_{\underline{12\dots n}}\right) \log \left(k_b-k_{\underline{12\dots n}}\right)-\log \left(k_a-k_{\underline{12\dots n}}\right) \log \left(k_b+k_{\underline{12\dots n}}\right)
\\
+\frac{1}{6} \left(3 \log ^2\left(k_a-k_{\underline{12\dots n}}\right)+3 \log ^2\left(k_b-k_{\underline{12\dots n}}\right)+\pi ^2\right)
\Bigg]
\end{multline}
Note that one still needs to integrate this result with respect to $\omega_1$ and $\omega_2$ by including appropriate $\tl\lambda(\omega_i)$ factors. However, we know that $\phi^4$ interaction in AdS$_4$ is actually conformally invariant so we do not need $\omega$ integrations if we focus on $\phi^4$ interaction. Indeed, we can directly write the full Witten expression for the second diagram in \figref{\ref{fig: loops}} as\footnote{This contradicts \equref{eq: truncated amplitude vs Witten diagram} because it is valid only for tree-level diagrams as we there make use of the fact that number of propagators is equal to number of vertices minus one.}
\be 
W=\lambda_4^2 \cA^L\left(k_1+k_2,k_3+k_4,\bm{k}_1+\bm{k}_2\right)
\ee 

\section{Conclusion}
In this paper, we have explored momentum space approach to computing amplitudes for scalars propagating in Anti-de Sitter space.  Adopting the algorithm provided in 
\cite{Arkani-Hamed:2017fdk} for cosmological wavefunctions, we compute both tree and loop level examples of AdS transition amplitudes. 

Our momentum space formalism provides a systematic and complementary study of scalars in AdS.
We are interested in using this formalism to computing higher point scalar loop amplitudes, which we leave to a future study. Likewise, this formalism can be utilized for computation of spinning loops. Unlike scalars, computing gluon and graviton loops  in general dimensions is complicated, but one can get nice results if one focuses on specific dimensions \cite{Albayrak:2020bso}.

There are several promising avenues for further explorations. We believe that there should be a natural polytopic interpretation to the results that we have developed, parallel to the interpretation in \cite{Arkani-Hamed:2017fdk}. It is also interesting whether the weight-shifting operators developed in \cite{Arkani-Hamed:2018kmz} in the context of cosmology can be generalized into our formalism, allowing us to relate spinning momentum space amplitudes to scalar momentum space amplitudes.\footnote{Such operators in positions space are constructed in \cite{Costa:2018mcg}.} Lastly, momentum space approach have been used to construct crossing symmetric basis for CFT correlators \cite{Isono:2019wex, Isono:2018rrb}. Our approach could be useful in such explorations as well. 

\section*{Acknowledgement}
SA is supported by DOE grant no.
DE-SC0020318 and Simons Foundation grant 488651. SK was supported by DRFC Discretionary
Funds from Williams College.
% END
\bibliographystyle{utphys}
\bibliography{references}{}

\providecommand{\href}[2]{#2}\begingroup\raggedright\begin{thebibliography}{10}

\bibitem{Maldacena:1997re}
J.~M. Maldacena, ``{The Large N limit of superconformal field theories and
  supergravity},'' \href{http://dx.doi.org/10.1023/A:1026654312961,
  10.4310/ATMP.1998.v2.n2.a1}{{\em Int. J. Theor. Phys.} {\bfseries 38} (1999)
  1113--1133}, \href{http://arxiv.org/abs/hep-th/9711200}{{\ttfamily
  arXiv:hep-th/9711200 [hep-th]}}.
[Adv. Theor. Math. Phys.2,231(1998)].
%%CITATION = HEP-TH/9711200;%%.

\bibitem{Witten:1998qj}
E.~Witten, ``{Anti-de Sitter space and holography},''
  \href{http://dx.doi.org/10.4310/ATMP.1998.v2.n2.a2}{{\em Adv. Theor. Math.
  Phys.} {\bfseries 2} (1998) 253--291},
\href{http://arxiv.org/abs/hep-th/9802150}{{\ttfamily arXiv:hep-th/9802150
  [hep-th]}}.
%%CITATION = HEP-TH/9802150;%%.

\bibitem{Gubser:1998bc}
S.~S. Gubser, I.~R. Klebanov, and A.~M. Polyakov, ``{Gauge theory correlators
  from noncritical string theory},''
  \href{http://dx.doi.org/10.1016/S0370-2693(98)00377-3}{{\em Phys. Lett.}
  {\bfseries B428} (1998) 105--114},
\href{http://arxiv.org/abs/hep-th/9802109}{{\ttfamily arXiv:hep-th/9802109
  [hep-th]}}.
%%CITATION = HEP-TH/9802109;%%.

\bibitem{Freedman:1998bj}
D.~Z. Freedman, S.~D. Mathur, A.~Matusis, and L.~Rastelli, ``{Comments on 4
  point functions in the CFT / AdS correspondence},''
  \href{http://dx.doi.org/10.1016/S0370-2693(99)00229-4}{{\em Phys. Lett.}
  {\bfseries B452} (1999) 61--68},
\href{http://arxiv.org/abs/hep-th/9808006}{{\ttfamily arXiv:hep-th/9808006
  [hep-th]}}.
%%CITATION = HEP-TH/9808006;%%.

\bibitem{Liu:1998ty}
H.~Liu and A.~A. Tseytlin, ``{On four point functions in the CFT / AdS
  correspondence},'' \href{http://dx.doi.org/10.1103/PhysRevD.59.086002}{{\em
  Phys. Rev.} {\bfseries D59} (1999) 086002},
\href{http://arxiv.org/abs/hep-th/9807097}{{\ttfamily arXiv:hep-th/9807097
  [hep-th]}}.
%%CITATION = HEP-TH/9807097;%%.

\bibitem{Freedman:1998tz}
D.~Z. Freedman, S.~D. Mathur, A.~Matusis, and L.~Rastelli, ``{Correlation
  functions in the CFT(d) / AdS(d+1) correspondence},''
  \href{http://dx.doi.org/10.1016/S0550-3213(99)00053-X}{{\em Nucl. Phys.}
  {\bfseries B546} (1999) 96--118},
\href{http://arxiv.org/abs/hep-th/9804058}{{\ttfamily arXiv:hep-th/9804058
  [hep-th]}}.
%%CITATION = HEP-TH/9804058;%%.

\bibitem{DHoker:1999mqo}
E.~D'Hoker, D.~Z. Freedman, and L.~Rastelli, ``{AdS / CFT four point functions:
  How to succeed at z integrals without really trying},''
  \href{http://dx.doi.org/10.1016/S0550-3213(99)00526-X}{{\em Nucl. Phys.}
  {\bfseries B562} (1999) 395--411},
\href{http://arxiv.org/abs/hep-th/9905049}{{\ttfamily arXiv:hep-th/9905049
  [hep-th]}}.
%%CITATION = HEP-TH/9905049;%%.

\bibitem{DHoker:1999kzh}
E.~D'Hoker, D.~Z. Freedman, S.~D. Mathur, A.~Matusis, and L.~Rastelli,
  ``{Graviton exchange and complete four point functions in the AdS / CFT
  correspondence},''
  \href{http://dx.doi.org/10.1016/S0550-3213(99)00525-8}{{\em Nucl. Phys.}
  {\bfseries B562} (1999) 353--394},
\href{http://arxiv.org/abs/hep-th/9903196}{{\ttfamily arXiv:hep-th/9903196
  [hep-th]}}.
%%CITATION = HEP-TH/9903196;%%.

\bibitem{Penedones:2010ue}
J.~Penedones, ``{Writing CFT correlation functions as AdS scattering
  amplitudes},'' \href{http://dx.doi.org/10.1007/JHEP03(2011)025}{{\em JHEP}
  {\bfseries 03} (2011) 025},
\href{http://arxiv.org/abs/1011.1485}{{\ttfamily arXiv:1011.1485 [hep-th]}}.
%%CITATION = ARXIV:1011.1485;%%.

\bibitem{Paulos:2011ie}
M.~F. Paulos, ``{Towards Feynman rules for Mellin amplitudes},''
  \href{http://dx.doi.org/10.1007/JHEP10(2011)074}{{\em JHEP} {\bfseries 10}
  (2011) 074},
\href{http://arxiv.org/abs/1107.1504}{{\ttfamily arXiv:1107.1504 [hep-th]}}.
%%CITATION = ARXIV:1107.1504;%%.

\bibitem{Mack:2009gy}
G.~Mack, ``{D-dimensional Conformal Field Theories with anomalous dimensions as
  Dual Resonance Models},'' {\em Bulg. J. Phys.} {\bfseries 36} (2009)
  214--226,
\href{http://arxiv.org/abs/0909.1024}{{\ttfamily arXiv:0909.1024 [hep-th]}}.
%%CITATION = ARXIV:0909.1024;%%.

\bibitem{Fitzpatrick:2011ia}
A.~L. Fitzpatrick, J.~Kaplan, J.~Penedones, S.~Raju, and B.~C. van Rees, ``{A
  Natural Language for AdS/CFT Correlators},''
  \href{http://dx.doi.org/10.1007/JHEP11(2011)095}{{\em JHEP} {\bfseries 11}
  (2011) 095},
\href{http://arxiv.org/abs/1107.1499}{{\ttfamily arXiv:1107.1499 [hep-th]}}.
%%CITATION = ARXIV:1107.1499;%%.

\bibitem{Kharel:2013mka}
S.~Kharel and G.~Siopsis, ``{Tree-level Correlators of scalar and vector fields
  in AdS/CFT},'' \href{http://dx.doi.org/10.1007/JHEP11(2013)159}{{\em JHEP}
  {\bfseries 11} (2013) 159},
\href{http://arxiv.org/abs/1308.2515}{{\ttfamily arXiv:1308.2515 [hep-th]}}.
%%CITATION = ARXIV:1308.2515;%%.

\bibitem{Fitzpatrick:2011hu}
A.~L. Fitzpatrick and J.~Kaplan, ``{Analyticity and the Holographic
  S-Matrix},'' \href{http://dx.doi.org/10.1007/JHEP10(2012)127}{{\em JHEP}
  {\bfseries 10} (2012) 127},
\href{http://arxiv.org/abs/1111.6972}{{\ttfamily arXiv:1111.6972 [hep-th]}}.
%%CITATION = ARXIV:1111.6972;%%.

\bibitem{Fitzpatrick:2011dm}
A.~L. Fitzpatrick and J.~Kaplan, ``{Unitarity and the Holographic S-Matrix},''
  \href{http://dx.doi.org/10.1007/JHEP10(2012)032}{{\em JHEP} {\bfseries 10}
  (2012) 032},
\href{http://arxiv.org/abs/1112.4845}{{\ttfamily arXiv:1112.4845 [hep-th]}}.
%%CITATION = ARXIV:1112.4845;%%.

\bibitem{Costa:2014kfa}
M.~S. Costa, V.~Gonalves, and J.~Penedones, ``{Spinning AdS Propagators},''
  \href{http://dx.doi.org/10.1007/JHEP09(2014)064}{{\em JHEP} {\bfseries 09}
  (2014) 064},
\href{http://arxiv.org/abs/1404.5625}{{\ttfamily arXiv:1404.5625 [hep-th]}}.
%%CITATION = ARXIV:1404.5625;%%.

\bibitem{Jepsen:2018ajn}
C.~B. Jepsen and S.~Parikh, ``{Recursion Relations in $p$-adic Mellin Space},''
  \href{http://dx.doi.org/10.1088/1751-8121/ab227b}{{\em J. Phys.} {\bfseries
  A52} no.~28, (2019) 285401},
\href{http://arxiv.org/abs/1812.09801}{{\ttfamily arXiv:1812.09801 [hep-th]}}.
%%CITATION = ARXIV:1812.09801;%%.

\bibitem{Jepsen:2018dqp}
C.~B. Jepsen and S.~Parikh, ``{$p$-adic Mellin Amplitudes},''
  \href{http://dx.doi.org/10.1007/JHEP04(2019)101}{{\em JHEP} {\bfseries 04}
  (2019) 101},
\href{http://arxiv.org/abs/1808.08333}{{\ttfamily arXiv:1808.08333 [hep-th]}}.
%%CITATION = ARXIV:1808.08333;%%.

\bibitem{Gubser:2018cha}
S.~S. Gubser, C.~Jepsen, and B.~Trundy, ``{Spin in $p$-adic AdS/CFT},''
  \href{http://dx.doi.org/10.1088/1751-8121/ab0757}{{\em J. Phys.} {\bfseries
  A52} no.~14, (2019) 144004},
\href{http://arxiv.org/abs/1811.02538}{{\ttfamily arXiv:1811.02538 [hep-th]}}.
%%CITATION = ARXIV:1811.02538;%%.

\bibitem{Hijano:2015zsa}
E.~Hijano, P.~Kraus, E.~Perlmutter, and R.~Snively, ``{Witten Diagrams
  Revisited: The AdS Geometry of Conformal Blocks},''
  \href{http://dx.doi.org/10.1007/JHEP01(2016)146}{{\em JHEP} {\bfseries 01}
  (2016) 146},
\href{http://arxiv.org/abs/1508.00501}{{\ttfamily arXiv:1508.00501 [hep-th]}}.
%%CITATION = ARXIV:1508.00501;%%.

\bibitem{Yuan:2018qva}
E.~Y. Yuan, ``{Simplicity in AdS Perturbative Dynamics},''
\href{http://arxiv.org/abs/1801.07283}{{\ttfamily arXiv:1801.07283 [hep-th]}}.
%%CITATION = ARXIV:1801.07283;%%.

\bibitem{Ghosh:2018bgd}
K.~Ghosh, ``{Polyakov-Mellin Bootstrap for AdS loops},''
\href{http://arxiv.org/abs/1811.00504}{{\ttfamily arXiv:1811.00504 [hep-th]}}.
%%CITATION = ARXIV:1811.00504;%%.

\bibitem{Liu:2018jhs}
J.~Liu, E.~Perlmutter, V.~Rosenhaus, and D.~Simmons-Duffin, ``{$d$-dimensional
  SYK, AdS Loops, and $6j$ Symbols},''
  \href{http://dx.doi.org/10.1007/JHEP03(2019)052}{{\em JHEP} {\bfseries 03}
  (2019) 052},
\href{http://arxiv.org/abs/1808.00612}{{\ttfamily arXiv:1808.00612 [hep-th]}}.
%%CITATION = ARXIV:1808.00612;%%.

\bibitem{Parikh:2019ygo}
S.~Parikh, ``{Holographic dual of the five-point conformal block},''
\href{http://arxiv.org/abs/1901.01267}{{\ttfamily arXiv:1901.01267 [hep-th]}}.
%%CITATION = ARXIV:1901.01267;%%.

\bibitem{Jepsen:2019svc}
C.~B. Jepsen and S.~Parikh, ``{Propagator identities, holographic conformal
  blocks, and higher-point AdS diagrams},''
  \href{http://dx.doi.org/10.1007/JHEP10(2019)268}{{\em JHEP} {\bfseries 10}
  (2019) 268},
\href{http://arxiv.org/abs/1906.08405}{{\ttfamily arXiv:1906.08405 [hep-th]}}.
%%CITATION = ARXIV:1906.08405;%%.

\bibitem{Penedones:2019tng}
J.~Penedones, J.~A. Silva, and A.~Zhiboedov, ``{Nonperturbative Mellin
  Amplitudes: Existence, Properties, Applications},''
\href{http://arxiv.org/abs/1912.11100}{{\ttfamily arXiv:1912.11100 [hep-th]}}.
%%CITATION = ARXIV:1912.11100;%%.

\bibitem{Aprile:2019rep}
F.~Aprile, J.~Drummond, P.~Heslop, and H.~Paul, ``{One-loop amplitudes in
  $AdS_5\times S^5$ supergravity from $\mathcal{N}=4$ SYM at strong
  coupling},''
\href{http://arxiv.org/abs/1912.01047}{{\ttfamily arXiv:1912.01047 [hep-th]}}.
%%CITATION = ARXIV:1912.01047;%%.

\bibitem{Ponomarev:2019ofr}
D.~Ponomarev, ``{From bulk loops to boundary large-N expansion},''
\href{http://arxiv.org/abs/1908.03974}{{\ttfamily arXiv:1908.03974 [hep-th]}}.
%%CITATION = ARXIV:1908.03974;%%.

\bibitem{Meltzer:2019pyl}
D.~Meltzer, ``{AdS/CFT Unitarity at Higher Loops: High-Energy String
  Scattering},''
\href{http://arxiv.org/abs/1912.05580}{{\ttfamily arXiv:1912.05580 [hep-th]}}.
%%CITATION = ARXIV:1912.05580;%%.

\bibitem{Meltzer:2019nbs}
D.~Meltzer, E.~Perlmutter, and A.~Sivaramakrishnan, ``{Unitarity Methods in
  AdS/CFT},''
\href{http://arxiv.org/abs/1912.09521}{{\ttfamily arXiv:1912.09521 [hep-th]}}.
%%CITATION = ARXIV:1912.09521;%%.

\bibitem{Caron-Huot:2018kta}
S.~Caron-Huot and A.-K. Trinh, ``{All tree-level correlators in
  AdS$_{5}$S$_{5}$ supergravity: hidden ten-dimensional conformal symmetry},''
  \href{http://dx.doi.org/10.1007/JHEP01(2019)196}{{\em JHEP} {\bfseries 01}
  (2019) 196},
\href{http://arxiv.org/abs/1809.09173}{{\ttfamily arXiv:1809.09173 [hep-th]}}.
%%CITATION = ARXIV:1809.09173;%%.

\bibitem{Fichet:2019hkg}
S.~Fichet, ``{Opacity and effective field theory in anti–de Sitter
  backgrounds},'' \href{http://dx.doi.org/10.1103/PhysRevD.100.095002}{{\em
  Phys. Rev.} {\bfseries D100} no.~9, (2019) 095002},
\href{http://arxiv.org/abs/1905.05779}{{\ttfamily arXiv:1905.05779 [hep-th]}}.
%%CITATION = ARXIV:1905.05779;%%.

\bibitem{Raju:2010by}
S.~Raju, ``{BCFW for Witten Diagrams},''
  \href{http://dx.doi.org/10.1103/PhysRevLett.106.091601}{{\em Phys. Rev.
  Lett.} {\bfseries 106} (2011) 091601},
\href{http://arxiv.org/abs/1011.0780}{{\ttfamily arXiv:1011.0780 [hep-th]}}.
%%CITATION = ARXIV:1011.0780;%%.

\bibitem{Mata:2012bx}
I.~Mata, S.~Raju, and S.~Trivedi, ``{CMB from CFT},''
  \href{http://dx.doi.org/10.1007/JHEP07(2013)015}{{\em JHEP} {\bfseries 07}
  (2013) 015},
\href{http://arxiv.org/abs/1211.5482}{{\ttfamily arXiv:1211.5482 [hep-th]}}.
%%CITATION = ARXIV:1211.5482;%%.

\bibitem{Raju:2012zs}
S.~Raju, ``{Four Point Functions of the Stress Tensor and Conserved Currents in
  AdS$_4$/CFT$_3$},'' \href{http://dx.doi.org/10.1103/PhysRevD.85.126008}{{\em
  Phys. Rev.} {\bfseries D85} (2012) 126008},
\href{http://arxiv.org/abs/1201.6452}{{\ttfamily arXiv:1201.6452 [hep-th]}}.
%%CITATION = ARXIV:1201.6452;%%.

\bibitem{Raju:2012zr}
S.~Raju, ``{New Recursion Relations and a Flat Space Limit for AdS/CFT
  Correlators},'' \href{http://dx.doi.org/10.1103/PhysRevD.85.126009}{{\em
  Phys. Rev.} {\bfseries D85} (2012) 126009},
\href{http://arxiv.org/abs/1201.6449}{{\ttfamily arXiv:1201.6449 [hep-th]}}.
%%CITATION = ARXIV:1201.6449;%%.

\bibitem{Raju:2011mp}
S.~Raju, ``{Recursion Relations for AdS/CFT Correlators},''
  \href{http://dx.doi.org/10.1103/PhysRevD.83.126002}{{\em Phys. Rev.}
  {\bfseries D83} (2011) 126002},
\href{http://arxiv.org/abs/1102.4724}{{\ttfamily arXiv:1102.4724 [hep-th]}}.
%%CITATION = ARXIV:1102.4724;%%.

\bibitem{Arkani-Hamed:2015bza}
N.~Arkani-Hamed and J.~Maldacena, ``{Cosmological Collider Physics},''
\href{http://arxiv.org/abs/1503.08043}{{\ttfamily arXiv:1503.08043 [hep-th]}}.
%%CITATION = ARXIV:1503.08043;%%.

\bibitem{Bzowski:2013sza}
A.~Bzowski, P.~McFadden, and K.~Skenderis, ``{Implications of conformal
  invariance in momentum space},''
  \href{http://dx.doi.org/10.1007/JHEP03(2014)111}{{\em JHEP} {\bfseries 03}
  (2014) 111},
\href{http://arxiv.org/abs/1304.7760}{{\ttfamily arXiv:1304.7760 [hep-th]}}.
%%CITATION = ARXIV:1304.7760;%%.

\bibitem{Bzowski:2015pba}
A.~Bzowski, P.~McFadden, and K.~Skenderis, ``{Scalar 3-point functions in CFT:
  renormalisation, beta functions and anomalies},''
  \href{http://dx.doi.org/10.1007/JHEP03(2016)066}{{\em JHEP} {\bfseries 03}
  (2016) 066},
\href{http://arxiv.org/abs/1510.08442}{{\ttfamily arXiv:1510.08442 [hep-th]}}.
%%CITATION = ARXIV:1510.08442;%%.

\bibitem{Bzowski:2018fql}
A.~Bzowski, P.~McFadden, and K.~Skenderis, ``{Renormalised CFT 3-point
  functions of scalars, currents and stress tensors},''
  \href{http://dx.doi.org/10.1007/JHEP11(2018)159}{{\em JHEP} {\bfseries 11}
  (2018) 159},
\href{http://arxiv.org/abs/1805.12100}{{\ttfamily arXiv:1805.12100 [hep-th]}}.
%%CITATION = ARXIV:1805.12100;%%.

\bibitem{Bzowski:2015yxv}
A.~Bzowski, P.~McFadden, and K.~Skenderis, ``{Evaluation of conformal
  integrals},'' \href{http://dx.doi.org/10.1007/JHEP02(2016)068}{{\em JHEP}
  {\bfseries 02} (2016) 068},
\href{http://arxiv.org/abs/1511.02357}{{\ttfamily arXiv:1511.02357 [hep-th]}}.
%%CITATION = ARXIV:1511.02357;%%.

\bibitem{Isono:2019ihz}
H.~Isono, T.~Noumi, and T.~Takeuchi, ``{Momentum space conformal three-point
  functions of conserved currents and a general spinning operator},''
  \href{http://dx.doi.org/10.1007/JHEP05(2019)057}{{\em JHEP} {\bfseries 05}
  (2019) 057},
\href{http://arxiv.org/abs/1903.01110}{{\ttfamily arXiv:1903.01110 [hep-th]}}.
%%CITATION = ARXIV:1903.01110;%%.

\bibitem{Isono:2018rrb}
H.~Isono, T.~Noumi, and G.~Shiu, ``{Momentum space approach to crossing
  symmetric CFT correlators},''
  \href{http://dx.doi.org/10.1007/JHEP07(2018)136}{{\em JHEP} {\bfseries 07}
  (2018) 136},
\href{http://arxiv.org/abs/1805.11107}{{\ttfamily arXiv:1805.11107 [hep-th]}}.
%%CITATION = ARXIV:1805.11107;%%.

\bibitem{Isono:2019wex}
H.~Isono, T.~Noumi, and G.~Shiu, ``{Momentum space approach to crossing
  symmetric CFT correlators. Part II. General spacetime dimension},''
  \href{http://dx.doi.org/10.1007/JHEP10(2019)183}{{\em JHEP} {\bfseries 10}
  (2019) 183},
\href{http://arxiv.org/abs/1908.04572}{{\ttfamily arXiv:1908.04572 [hep-th]}}.
%%CITATION = ARXIV:1908.04572;%%.

\bibitem{Coriano:2018bbe}
C.~Corian{\`o} and M.~M. Maglio, ``{Exact Correlators from Conformal Ward
  Identities in Momentum Space and the Perturbative $TJJ$ Vertex},''
  \href{http://dx.doi.org/10.1016/j.nuclphysb.2018.11.016}{{\em Nucl. Phys.}
  {\bfseries B938} (2019) 440--522},
\href{http://arxiv.org/abs/1802.07675}{{\ttfamily arXiv:1802.07675 [hep-th]}}.
%%CITATION = ARXIV:1802.07675;%%.

\bibitem{Coriano:2019dyc}
C.~Corian{\`o}, M.~M. Maglio, A.~Tatullo, and D.~Theofilopoulos, ``{Exact
  Correlators from Conformal Ward Identities in Momentum Space and Perturbative
  Realizations},'' in {\em {18th Hellenic School and Workshops on Elementary
  Particle Physics and Gravity (CORFU2018) Corfu, Corfu, Greece, August
  31-September 28, 2018}}.
\newblock 2019.
\newblock
\href{http://arxiv.org/abs/1904.13174}{{\ttfamily arXiv:1904.13174 [hep-ph]}}.
\newblock
%%CITATION = ARXIV:1904.13174;%%.

\bibitem{Maglio:2019grh}
C.~Corian{\`o} and M.~M. Maglio, ``{On Some Hypergeometric Solutions of the
  Conformal Ward Identities of Scalar 4-point Functions in Momentum Space},''
\href{http://arxiv.org/abs/1903.05047}{{\ttfamily arXiv:1903.05047 [hep-th]}}.
%%CITATION = ARXIV:1903.05047;%%.

\bibitem{Gillioz:2018mto}
M.~Gillioz, ``{Momentum-space conformal blocks on the light cone},''
  \href{http://dx.doi.org/10.1007/JHEP10(2018)125}{{\em JHEP} {\bfseries 10}
  (2018) 125},
\href{http://arxiv.org/abs/1807.07003}{{\ttfamily arXiv:1807.07003 [hep-th]}}.
%%CITATION = ARXIV:1807.07003;%%.

\bibitem{Arkani-Hamed:2018kmz}
N.~Arkani-Hamed, D.~Baumann, H.~Lee, and G.~L. Pimentel, ``{The Cosmological
  Bootstrap: Inflationary Correlators from Symmetries and Singularities},''
\href{http://arxiv.org/abs/1811.00024}{{\ttfamily arXiv:1811.00024 [hep-th]}}.
%%CITATION = ARXIV:1811.00024;%%.

\bibitem{Coriano:2018bsy}
C.~Corian{\`o} and M.~M. Maglio, ``{The general 3-graviton vertex ($TTT$) of
  conformal field theories in momentum space in $d = 4$},''
  \href{http://dx.doi.org/10.1016/j.nuclphysb.2018.10.007}{{\em Nucl. Phys.}
  {\bfseries B937} (2018) 56--134},
\href{http://arxiv.org/abs/1808.10221}{{\ttfamily arXiv:1808.10221 [hep-th]}}.
%%CITATION = ARXIV:1808.10221;%%.

\bibitem{Coriano:2013jba}
C.~Corian{\`o}, L.~Delle~Rose, E.~Mottola, and M.~Serino, ``{Solving the
  Conformal Constraints for Scalar Operators in Momentum Space and the
  Evaluation of Feynman's Master Integrals},''
  \href{http://dx.doi.org/10.1007/JHEP07(2013)011}{{\em JHEP} {\bfseries 07}
  (2013) 011},
\href{http://arxiv.org/abs/1304.6944}{{\ttfamily arXiv:1304.6944 [hep-th]}}.
%%CITATION = ARXIV:1304.6944;%%.

\bibitem{Bzowski:2019kwd}
A.~Bzowski, P.~McFadden, and K.~Skenderis, ``{Conformal $n$-point functions in
  momentum space},''
\href{http://arxiv.org/abs/1910.10162}{{\ttfamily arXiv:1910.10162 [hep-th]}}.
%%CITATION = ARXIV:1910.10162;%%.

\bibitem{Anand:2019lkt}
N.~Anand, Z.~U. Khandker, and M.~T. Walters, ``{Momentum space CFT correlators
  for Hamiltonian truncation},''
\href{http://arxiv.org/abs/1911.02573}{{\ttfamily arXiv:1911.02573 [hep-th]}}.
%%CITATION = ARXIV:1911.02573;%%.

\bibitem{Gillioz:2019lgs}
M.~Gillioz, ``{Conformal 3-point functions and the Lorentzian OPE in momentum
  space},''
\href{http://arxiv.org/abs/1909.00878}{{\ttfamily arXiv:1909.00878 [hep-th]}}.
%%CITATION = ARXIV:1909.00878;%%.

\bibitem{Farrow:2018yni}
J.~A. Farrow, A.~E. Lipstein, and P.~McFadden, ``{Double copy structure of CFT
  correlators},'' \href{http://dx.doi.org/10.1007/JHEP02(2019)130}{{\em JHEP}
  {\bfseries 02} (2019) 130},
\href{http://arxiv.org/abs/1812.11129}{{\ttfamily arXiv:1812.11129 [hep-th]}}.
%%CITATION = ARXIV:1812.11129;%%.

\bibitem{Lipstein:2019mpu}
A.~Lipstein and P.~McFadden, ``{Double copy structure and the flat space limit
  of conformal correlators in even dimensions},''
\href{http://arxiv.org/abs/1912.10046}{{\ttfamily arXiv:1912.10046 [hep-th]}}.
%%CITATION = ARXIV:1912.10046;%%.

\bibitem{Nagaraj:2019zmk}
B.~Nagaraj and D.~Ponomarev, ``{Spinor-Helicity Formalism for Massless Fields
  in AdS$_4$ II: Potentials},''
\href{http://arxiv.org/abs/1912.07494}{{\ttfamily arXiv:1912.07494 [hep-th]}}.
%%CITATION = ARXIV:1912.07494;%%.

\bibitem{Albayrak:2018tam}
S.~Albayrak and S.~Kharel, ``{Towards the higher point holographic momentum
  space amplitudes},'' \href{http://dx.doi.org/10.1007/JHEP02(2019)040}{{\em
  JHEP} {\bfseries 02} (2019) 040},
\href{http://arxiv.org/abs/1810.12459}{{\ttfamily arXiv:1810.12459 [hep-th]}}.
%%CITATION = ARXIV:1810.12459;%%.

\bibitem{Albayrak:2019yve}
S.~Albayrak and S.~Kharel, ``{Towards the higher point holographic momentum
  space amplitudes II: Gravitons},''
  \href{http://dx.doi.org/10.1007/JHEP12(2019)135}{{\em JHEP} {\bfseries 12}
  (2019) 135},
\href{http://arxiv.org/abs/1908.01835}{{\ttfamily arXiv:1908.01835 [hep-th]}}.
%%CITATION = ARXIV:1908.01835;%%.

\bibitem{Arkani-Hamed:2017fdk}
N.~Arkani-Hamed, P.~Benincasa, and A.~Postnikov, ``{Cosmological Polytopes and
  the Wavefunction of the Universe},''
\href{http://arxiv.org/abs/1709.02813}{{\ttfamily arXiv:1709.02813 [hep-th]}}.
%%CITATION = ARXIV:1709.02813;%%.

\bibitem{Albayrak:2019asr}
S.~Albayrak, C.~Chowdhury, and S.~Kharel, ``{New relation for Witten
  diagrams},'' \href{http://dx.doi.org/10.1007/JHEP10(2019)274}{{\em JHEP}
  {\bfseries 10} (2019) 274},
\href{http://arxiv.org/abs/1904.10043}{{\ttfamily arXiv:1904.10043 [hep-th]}}.
%%CITATION = ARXIV:1904.10043;%%.

\bibitem{Maldacena:2002vr}
J.~M. Maldacena, ``{Non-Gaussian features of primordial fluctuations in single
  field inflationary models},''
  \href{http://dx.doi.org/10.1088/1126-6708/2003/05/013}{{\em JHEP} {\bfseries
  05} (2003) 013},
\href{http://arxiv.org/abs/astro-ph/0210603}{{\ttfamily arXiv:astro-ph/0210603
  [astro-ph]}}.
%%CITATION = ASTRO-PH/0210603;%%.

\bibitem{Ghosh:2014kba}
A.~Ghosh, N.~Kundu, S.~Raju, and S.~P. Trivedi, ``{Conformal Invariance and the
  Four Point Scalar Correlator in Slow-Roll Inflation},''
  \href{http://dx.doi.org/10.1007/JHEP07(2014)011}{{\em JHEP} {\bfseries 07}
  (2014) 011},
\href{http://arxiv.org/abs/1401.1426}{{\ttfamily arXiv:1401.1426 [hep-th]}}.
%%CITATION = ARXIV:1401.1426;%%.

\bibitem{McFadden:2009fg}
P.~McFadden and K.~Skenderis, ``{Holography for Cosmology},''
  \href{http://dx.doi.org/10.1103/PhysRevD.81.021301}{{\em Phys. Rev.}
  {\bfseries D81} (2010) 021301},
\href{http://arxiv.org/abs/0907.5542}{{\ttfamily arXiv:0907.5542 [hep-th]}}.
%%CITATION = ARXIV:0907.5542;%%.

\bibitem{Chen:2009zp}
X.~Chen and Y.~Wang, ``{Quasi-Single Field Inflation and Non-Gaussianities},''
  \href{http://dx.doi.org/10.1088/1475-7516/2010/04/027}{{\em JCAP} {\bfseries
  1004} (2010) 027},
\href{http://arxiv.org/abs/0911.3380}{{\ttfamily arXiv:0911.3380 [hep-th]}}.
%%CITATION = ARXIV:0911.3380;%%.

\bibitem{Maldacena:2011nz}
J.~M. Maldacena and G.~L. Pimentel, ``{On graviton non-Gaussianities during
  inflation},'' \href{http://dx.doi.org/10.1007/JHEP09(2011)045}{{\em JHEP}
  {\bfseries 09} (2011) 045},
\href{http://arxiv.org/abs/1104.2846}{{\ttfamily arXiv:1104.2846 [hep-th]}}.
%%CITATION = ARXIV:1104.2846;%%.

\bibitem{Baumann:2011nk}
D.~Baumann and D.~Green, ``{Signatures of Supersymmetry from the Early
  Universe},'' \href{http://dx.doi.org/10.1103/PhysRevD.85.103520}{{\em Phys.
  Rev.} {\bfseries D85} (2012) 103520},
\href{http://arxiv.org/abs/1109.0292}{{\ttfamily arXiv:1109.0292 [hep-th]}}.
%%CITATION = ARXIV:1109.0292;%%.

\bibitem{Assassi:2012zq}
V.~Assassi, D.~Baumann, and D.~Green, ``{On Soft Limits of Inflationary
  Correlation Functions},''
  \href{http://dx.doi.org/10.1088/1475-7516/2012/11/047}{{\em JCAP} {\bfseries
  1211} (2012) 047},
\href{http://arxiv.org/abs/1204.4207}{{\ttfamily arXiv:1204.4207 [hep-th]}}.
%%CITATION = ARXIV:1204.4207;%%.

\bibitem{Chen:2012ge}
X.~Chen and Y.~Wang, ``{Quasi-Single Field Inflation with Large Mass},''
  \href{http://dx.doi.org/10.1088/1475-7516/2012/09/021}{{\em JCAP} {\bfseries
  1209} (2012) 021},
\href{http://arxiv.org/abs/1205.0160}{{\ttfamily arXiv:1205.0160 [hep-th]}}.
%%CITATION = ARXIV:1205.0160;%%.

\bibitem{Assassi:2013gxa}
V.~Assassi, D.~Baumann, D.~Green, and L.~McAllister, ``{Planck-Suppressed
  Operators},'' \href{http://dx.doi.org/10.1088/1475-7516/2014/01/033}{{\em
  JCAP} {\bfseries 1401} (2014) 033},
\href{http://arxiv.org/abs/1304.5226}{{\ttfamily arXiv:1304.5226 [hep-th]}}.
%%CITATION = ARXIV:1304.5226;%%.

\bibitem{Lee:2016vti}
H.~Lee, D.~Baumann, and G.~L. Pimentel, ``{Non-Gaussianity as a Particle
  Detector},'' \href{http://dx.doi.org/10.1007/JHEP12(2016)040}{{\em JHEP}
  {\bfseries 12} (2016) 040},
\href{http://arxiv.org/abs/1607.03735}{{\ttfamily arXiv:1607.03735 [hep-th]}}.
%%CITATION = ARXIV:1607.03735;%%.

\bibitem{An:2017hlx}
H.~An, M.~McAneny, A.~K. Ridgway, and M.~B. Wise, ``{Quasi Single Field
  Inflation in the non-perturbative regime},''
  \href{http://dx.doi.org/10.1007/JHEP06(2018)105}{{\em JHEP} {\bfseries 06}
  (2018) 105},
\href{http://arxiv.org/abs/1706.09971}{{\ttfamily arXiv:1706.09971 [hep-ph]}}.
%%CITATION = ARXIV:1706.09971;%%.

\bibitem{Kehagias:2017cym}
A.~Kehagias and A.~Riotto, ``{On the Inflationary Perturbations of Massive
  Higher-Spin Fields},''
  \href{http://dx.doi.org/10.1088/1475-7516/2017/07/046}{{\em JCAP} {\bfseries
  1707} no.~07, (2017) 046},
\href{http://arxiv.org/abs/1705.05834}{{\ttfamily arXiv:1705.05834 [hep-th]}}.
%%CITATION = ARXIV:1705.05834;%%.

\bibitem{Kumar:2017ecc}
S.~Kumar and R.~Sundrum, ``{Heavy-Lifting of Gauge Theories By Cosmic
  Inflation},'' \href{http://dx.doi.org/10.1007/JHEP05(2018)011}{{\em JHEP}
  {\bfseries 05} (2018) 011},
\href{http://arxiv.org/abs/1711.03988}{{\ttfamily arXiv:1711.03988 [hep-ph]}}.
%%CITATION = ARXIV:1711.03988;%%.

\bibitem{Baumann:2017jvh}
D.~Baumann, G.~Goon, H.~Lee, and G.~L. Pimentel, ``{Partially Massless Fields
  During Inflation},'' \href{http://dx.doi.org/10.1007/JHEP04(2018)140}{{\em
  JHEP} {\bfseries 04} (2018) 140},
\href{http://arxiv.org/abs/1712.06624}{{\ttfamily arXiv:1712.06624 [hep-th]}}.
%%CITATION = ARXIV:1712.06624;%%.

\bibitem{Franciolini:2017ktv}
G.~Franciolini, A.~Kehagias, and A.~Riotto, ``{Imprints of Spinning Particles
  on Primordial Cosmological Perturbations},''
  \href{http://dx.doi.org/10.1088/1475-7516/2018/02/023}{{\em JCAP} {\bfseries
  1802} no.~02, (2018) 023},
\href{http://arxiv.org/abs/1712.06626}{{\ttfamily arXiv:1712.06626 [hep-th]}}.
%%CITATION = ARXIV:1712.06626;%%.

\bibitem{Goon:2018fyu}
G.~Goon, K.~Hinterbichler, A.~Joyce, and M.~Trodden, ``{Shapes of gravity:
  Tensor non-Gaussianity and massive spin-2 fields},''
\href{http://arxiv.org/abs/1812.07571}{{\ttfamily arXiv:1812.07571 [hep-th]}}.
%%CITATION = ARXIV:1812.07571;%%.

\bibitem{Anninos:2019nib}
D.~Anninos, V.~De~Luca, G.~Franciolini, A.~Kehagias, and A.~Riotto,
  ``{Cosmological Shapes of Higher-Spin Gravity},''
  \href{http://dx.doi.org/10.1088/1475-7516/2019/04/045}{{\em JCAP} {\bfseries
  1904} no.~04, (2019) 045},
\href{http://arxiv.org/abs/1902.01251}{{\ttfamily arXiv:1902.01251 [hep-th]}}.
%%CITATION = ARXIV:1902.01251;%%.

\bibitem{Pi:2012gf}
S.~Pi and M.~Sasaki, ``{Curvature Perturbation Spectrum in Two-field Inflation
  with a Turning Trajectory},''
  \href{http://dx.doi.org/10.1088/1475-7516/2012/10/051}{{\em JCAP} {\bfseries
  1210} (2012) 051},
\href{http://arxiv.org/abs/1205.0161}{{\ttfamily arXiv:1205.0161 [hep-th]}}.
%%CITATION = ARXIV:1205.0161;%%.

\bibitem{Gong:2013sma}
J.-O. Gong, S.~Pi, and M.~Sasaki, ``{Equilateral non-Gaussianity from heavy
  fields},'' \href{http://dx.doi.org/10.1088/1475-7516/2013/11/043}{{\em JCAP}
  {\bfseries 1311} (2013) 043},
\href{http://arxiv.org/abs/1306.3691}{{\ttfamily arXiv:1306.3691 [hep-th]}}.
%%CITATION = ARXIV:1306.3691;%%.

\bibitem{Sleight:2019mgd}
C.~Sleight, ``{A Mellin Space Approach to Cosmological Correlators},''
\href{http://arxiv.org/abs/1906.12302}{{\ttfamily arXiv:1906.12302 [hep-th]}}.
%%CITATION = ARXIV:1906.12302;%%.

\bibitem{Sleight:2019hfp}
C.~Sleight and M.~Taronna, ``{Bootstrapping Inflationary Correlators in Mellin
  Space},''
\href{http://arxiv.org/abs/1907.01143}{{\ttfamily arXiv:1907.01143 [hep-th]}}.
%%CITATION = ARXIV:1907.01143;%%.

\bibitem{Hillman:2019wgh}
A.~Hillman, ``{Symbol Recursion for the dS Wave Function},''
\href{http://arxiv.org/abs/1912.09450}{{\ttfamily arXiv:1912.09450 [hep-th]}}.
%%CITATION = ARXIV:1912.09450;%%.

\bibitem{Sonego:1993fw}
S.~Sonego and V.~Faraoni, ``{Coupling to the curvature for a scalar field from
  the equivalence principle},''
\href{http://dx.doi.org/10.1088/0264-9381/10/6/015}{{\em Class. Quant. Grav.}
  {\bfseries 10} (1993) 1185--1187}.
%%CITATION = CQGRD,10,1185;%%.

\bibitem{Birrell:1982ix}
N.~D. Birrell and P.~C.~W. Davies,
  \href{http://dx.doi.org/10.1017/CBO9780511622632}{{\em {Quantum Fields in
  Curved Space}}}.
\newblock Cambridge Monographs on Mathematical Physics. Cambridge Univ. Press,
  Cambridge, UK, 1984.
\newblock
\url{http://www.cambridge.org/mw/academic/subjects/physics/theoretical-physics-and-mathematical-physics/quantum-fields-curved-space?format=PB}.
\newblock
%%CITATION = INSPIRE-181166;%%.

\bibitem{Gibbons:1976ue}
G.~W. Gibbons and S.~W. Hawking, ``{Action Integrals and Partition Functions in
  Quantum Gravity},''
\href{http://dx.doi.org/10.1103/PhysRevD.15.2752}{{\em Phys. Rev.} {\bfseries
  D15} (1977) 2752--2756}.
%%CITATION = PHRVA,D15,2752;%%.

\bibitem{Guarnizo:2010xr}
A.~Guarnizo, L.~Castaneda, and J.~M. Tejeiro, ``{Boundary Term in Metric f(R)
  Gravity: Field Equations in the Metric Formalism},''
  \href{http://dx.doi.org/10.1007/s10714-010-1012-6}{{\em Gen. Rel. Grav.}
  {\bfseries 42} (2010) 2713--2728},
\href{http://arxiv.org/abs/1002.0617}{{\ttfamily arXiv:1002.0617 [gr-qc]}}.
%%CITATION = ARXIV:1002.0617;%%.

\bibitem{Nojiri:2000gv}
S.~Nojiri and S.~D. Odintsov, ``{Brane world cosmology in higher derivative
  gravity or warped compactification in the next-to-leading order of AdS / CFT
  correspondence},''
  \href{http://dx.doi.org/10.1088/1126-6708/2000/07/049}{{\em JHEP} {\bfseries
  07} (2000) 049},
\href{http://arxiv.org/abs/hep-th/0006232}{{\ttfamily arXiv:hep-th/0006232
  [hep-th]}}.
%%CITATION = HEP-TH/0006232;%%.

\bibitem{Carmi:2019ocp}
D.~Carmi, ``{Loops in AdS: From the Spectral Representation to Position
  Space},''
\href{http://arxiv.org/abs/1910.14340}{{\ttfamily arXiv:1910.14340 [hep-th]}}.
%%CITATION = ARXIV:1910.14340;%%.

\bibitem{Baumann:2019oyu}
D.~Baumann, C.~Duaso~Pueyo, A.~Joyce, H.~Lee, and G.~L. Pimentel, ``{The
  Cosmological Bootstrap: Weight-Shifting Operators and Scalar Seeds},''
\href{http://arxiv.org/abs/1910.14051}{{\ttfamily arXiv:1910.14051 [hep-th]}}.
%%CITATION = ARXIV:1910.14051;%%.

\bibitem{Faraoni:2013igs}
V.~Faraoni, ``{Conformally coupled inflation},''
\href{http://dx.doi.org/10.3390/galaxies1020096}{{\em Galaxies} {\bfseries 1}
  no.~2, (2013) 96--106}.
%%CITATION = INSPIRE-1250375;%%.

\bibitem{Giombi:2017hpr}
S.~Giombi, C.~Sleight, and M.~Taronna, ``{Spinning AdS Loop Diagrams: Two Point
  Functions},'' \href{http://dx.doi.org/10.1007/JHEP06(2018)030}{{\em JHEP}
  {\bfseries 06} (2018) 030},
\href{http://arxiv.org/abs/1708.08404}{{\ttfamily arXiv:1708.08404 [hep-th]}}.
%%CITATION = ARXIV:1708.08404;%%.

\bibitem{Albayrak:2020bso}
S.~Albayrak and S.~Kharel, ``{On spinning loop amplitudes in Anti-de Sitter
  space},'' \href{http://arxiv.org/abs/2006.12540}{{\ttfamily arXiv:2006.12540
  [hep-th]}}.

\bibitem{Costa:2018mcg}
M.~S. Costa and T.~Hansen, ``{AdS Weight Shifting Operators},''
\href{http://arxiv.org/abs/1805.01492}{{\ttfamily arXiv:1805.01492 [hep-th]}}.
%%CITATION = ARXIV:1805.01492;%%.

\end{thebibliography}\endgroup


\providecommand{\href}[2]{#2}\begingroup\raggedright\endgroup
\end{document}